\documentclass[numbers,sort&compress]{article}
\usepackage[utf8]{inputenc}
\usepackage[T1]{fontenc}
\usepackage{graphicx}
\usepackage{longtable}
\usepackage{wrapfig}
\usepackage{rotating}
\usepackage[normalem]{ulem}
\usepackage{amsmath}
\usepackage{amssymb}
\usepackage{capt-of}
\usepackage{hyperref}
\usepackage[frozencache]{minted}
\usepackage{arxiv}
\usepackage{amssymb,amsmath,amsthm}
\usepackage{natbib}
\usepackage{xcolor}
\usepackage{hyperref}
\usepackage{tabularx}
\usepackage[toc,page]{appendix}
\graphicspath{{Artworks/}}
\graphicspath{{imgs/}}
\usepackage{algorithm2e}
\theoremstyle{definition}
\usepackage{url}            
\usepackage{amsfonts}       
\usepackage{nicefrac}       
\usepackage{microtype}      
\usepackage{cleveref}       
\usepackage{graphicx}
\usepackage{doi}
\author{ \href{https://orcid.org/0000-0002-2393-8056}{\includegraphics[scale=0.06]{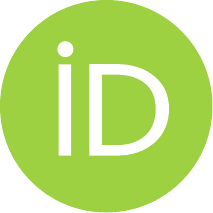}\hspace{1mm}Rohit Goswami}\thanks{Corresponding Author} \\
  Institute IMX and Lab-COSMO \\
  \'Ecole polytechnique f\'ed\'erale de Lausanne (EPFL) \\
  Station 12, CH-1015 Lausanne, Switzerland \\
  and \\
  Science Institute \& Faculty of Physical Sciences\\
  University of Iceland, 107 Reykjav\'ik, Iceland\\[1ex]
  \texttt{rgoswami@ieee.org}
  \And
  \href{https://orcid.org/0000-0001-8285-5421}{\includegraphics[scale=0.06]{orcid.pdf}\hspace{1mm}Hannes J\'onsson}\thanks{Corresponding Author} \\
  Science Institute \& Faculty of Physical Sciences\\
  University of Iceland\\
  107 Reykjav\'ik, Iceland\\
  \texttt{hj@hi.is}
}

\date{\today}
\title{Adaptive Pruning for Increased Robustness and Reduced Computational Overhead in Gaussian Process Accelerated Saddle Point Searches}
\hypersetup{
 pdfauthor={},
 pdftitle={Adaptive Pruning for Increased Robustness and Reduced Computational Overhead in Gaussian Process Accelerated Saddle Point Searches},
 pdfkeywords={},
 pdfsubject={},
 pdfcreator={Emacs 30.2 (Org mode 9.7.34)}, 
 pdflang={English}}
\begin{document}

\maketitle
\begin{abstract} 
Gaussian process (GP) regression provides a strategy for accelerating saddle point searches on high-dimensional energy surfaces by reducing the number of times the energy and its derivatives with respect to atomic coordinates need to be evaluated. The computational overhead in the hyperparameter optimization can, however, be large and make the approach inefficient.
Failures can also occur if the search ventures too far into regions that are not represented well enough by the GP model.
Here, these challenges are resolved by using geometry-aware optimal transport measures and an active pruning strategy using a summation over Wasserstein-1 distances for each atom-type in farthest-point sampling, selecting a fixed-size subset of geometrically diverse configurations to
avoid rapidly increasing cost of GP updates as more observations are made.
Stability is enhanced by permutation-invariant metric that provides a reliable trust radius for early-stopping and a logarithmic barrier penalty for the growth of the signal variance. These physically motivated algorithmic changes prove their efficacy by
reducing to less than a half the mean computational time on a set of 238 challenging configurations from a previously published data set of chemical reactions. With these improvements, the GP approach is established as
a robust and scalable algorithm for accelerating saddle point searches when the evaluation of the energy and atomic forces requires significant computational effort.
\end{abstract}
\keywords{Saddle Point Search, Gaussian Process Regression, Dimer Method, Optimal Transport, Data Pruning, Computational Chemistry}
\section{Introduction}
\label{sec:introduction}
An understanding of the transition mechanisms and estimation of the rate of atomic rearrangements, such as chemical reactions and diffusion events, requires analyzing the high-dimensional potential energy surface (PES) and identifying transition states.
The PES specifies how the energy of the system varies as a function of the atomic coordinates and the negative gradient gives the force acting on the atoms.
Reactant and product states correspond to local minima on the PES.
Within the harmonic approximation of transition state theory, first order saddle points on the PES, i.e. stationary points where the Hessian matrix has one negative eigenvalue, characterize the transition states.\cite{petersReactionRateTheory2017}
Similar considerations apply to magnetic transitions where magnetic vectors rotate.\cite{schrautzerIdentificationMechanismsMagnetic2025}

Several methods have been developed for finding first order saddle points on energy surfaces given only a starting configuration of the atoms \cite{cerjanFindingTransitionStates1981,walesFindingSaddlePoints1989,mousseauTravelingPotentialEnergy1998,henkelmanDimerMethodFinding1999,jayActivationRelaxationTechnique2022,munroDefectMigrationCrystalline1999,schlegelGeometryOptimization2011,hermesSellaOpenSourceAutomationFriendly2022} without a predefined final state of the transition. These are distinct from the ``double ended'' methods like the Nudged elastic band \cite{jonssonNudgedElasticBand1998,petersReactionRateTheory2017,henkelmanClimbingImageNudged2000,goswamiEfficientExplorationChemical2025} and variants thereof \cite{asgeirssonNudgedElasticBand2021} which require known initial and final states. We focus on general-purpose methods which do not require user-defined constraints on the transition state optimization \cite{fdez.galvanRestrictedVarianceConstrainedReaction2021} and compare to the extant state of the art within such methods \cite{goswamiEfficientImplementationGaussian2025a}.
A given initial state, without reference to a known product state, can in general lead to many different saddle points.
By making a catalog of the lower energy saddle points, which correspond to the faster transitions,
long time scale simulations of the system can be carried out for a given temperature.
\cite{henkelmanLongTimeScale2001,belandKineticActivationrelaxationTechnique2011}
But, the computational effort involved when the PES is obtained from an electronic structure calculation
makes this intractable.
General purpose machine learned interatomic potentials, such as PET-MAD, \cite{mazitovPETMADUniversalInteratomic2025,bigiMetatensorMetatomicFoundational2025} can alleviate some of this cost for systems
where a large set of electronic structure calculations has been performed, but these often suffer from under-sampling of the regions near the transition states as they are of low probability and are harder to sample.\cite{schreinerTransition1xDatasetBuilding2022}.

To improve efficiency, a local surrogate model of the PES can be created on-the-fly, most successfully using Gaussian Processes (GPs) \cite{koistinenMinimumModeSaddle2020,goswamiEfficientImplementationGaussian2025a,denzelGaussianProcessRegression2018a}, or neural networks \cite{petersonAccelerationSaddlepointSearches2016}.
This, however, introduces a second, often overlooked cost: the scaling of the computational effort in generating and updating the surrogate model. For GPs, the hyperparameter optimization involves repeatedly inverting a covariance matrix whose size grows with both the number of collected data points, \(M_{data}\), and the number of atoms, \(N_{atoms}\). \cite{kochenderferAlgorithmsOptimization,rasmussenGaussianProcessesMachine2006,gramacySurrogatesGaussianProcess2020}
This steep computational scaling, especially when the gradient of the energy, i.e. atomic forces, are included in the training, can cause the wall-time cost of the surrogate model to eventually exceed that of the electronic structure calculations, creating a new performance bottleneck. It is, therefore, important to develop ways to generate accurate enough surrogate models with lower computational overhead.

Here, a framework for generating a surrogate GP energy surface is introduced where the scaling of the computational overhead remains nearly constant as additional data is acquired. Several features are, furthermore, introduced to make the construction and the use of the surrogate model more robust. We refer to the method as Optimal Transport Gaussian Process (OT-GP). Only a representative subset of the data is used to fit the hyperparamenters in the GP model, selected by a Farthest Point Sampling (FPS) strategy with the Earth Mover's Distance (EMD) measure. Several stability enhancements are also introduced, including a data-driven adaptive early stopping threshold to balance safety in exploration, a hyperparameter oscillation detection heuristic to ensure stable learning, an adaptive variance barrier to prevent unphysical jumps in the predicted PES, and a rigorous removal of overal rotation of the system.
This geometry-aware subsampling and stabilization unlocks true wall-time efficiency for the GP accelerated saddle searches.

In Section \ref{sec:cmeth}, the methodology is described, including a review of the previous
state-of-the-art in \ref{sec:cmeth:background}, a presentation of the OT-GP algorithm in \ref{sec:cmeth:otgp} with computational specifics for the systems studied given in \ref{sec:cmeth:cdet}.
Section \ref{sec:results} presents results of saddle point searches for molecular systems in terms of reliability, quality of saddles found, and performance metrics, using illustrative examples to validate the algorithmic choices and a large-scale benchmark test.
Section \ref{sec:discussion} provides discussions on the physical and data-driven insights, analyzing the algorithm and the interpretation of the hyperparameters.
Conclusions are given in Section \ref{sec:conclusion} as well as indications of future directions, highlighting the method's robustness and long-term software maintenance.
\section{Computational Methods}
\label{sec:cmeth}
\subsection{Dimer Method and GP Acceleration}
\label{sec:cmeth:background}
The dimer method and its variants have been described in several articles  \cite{henkelmanDimerMethodFinding1999,olsenComparisonMethodsFinding2004,heydenEfficientMethodsFinding2005,kastnerSuperlinearlyConvergingDimer2008,pedersenEfficientSamplingSaddle2011,lengEfficientSoftestMode2013,goswamiBayesianHierarchicalModels2025a,goswamiEfficientImplementationGaussian2025a}. Here, we recap the standard formulation as implemented in the EON software \cite{chillEONSoftwareLong2014}
\footnote{Documentation available at \url{https://eondocs.org}}.
The dimer method is a minimum mode following approach where the eigenvector corresponding to the minimum eigenvalue of the Hessian,
the matrix of second derivatives of \(V(\mathbb{R})\) with respect to atom coordinates,
is found by rotating two copies of the system, the dimer, so as to find the minimum in total energy for a given location of its center.
The component of the gradient of the energy \(V(\mathbb{R})\) in the direction of the minimum mode is then inverted and the dimer
displaced opposite to the direction of the modified gradient. This leads in an iterative way to convergence on a first order saddle point.
Let \(x \in \mathbb{R}^{3n}\) denote the position vector in the extended configuration space of all the movable atoms \(\mathbb{R}^{n\times3}\to\mathbb{R}^{3n}\).
The iterative path towards the saddle point involves moving uphill in energy along the minimum mode while going downhill in all orthogonal
directions
\begin{align}
\dot x &= \mathbf{F}^\perp - \mathbf{F}^\parallel \\
\mathbf{F}^\parallel &= -\hat\tau^T\nabla V(x)\hat\tau \\
\mathbf{F}^\perp &= -\nabla V(x) - \mathbf{F}^\parallel
\label{eq:dimer_base}
\end{align}
where \(\hat\tau\) represents the unit eigenvector corresponding to the lowest eigenvalue of the Hessian matrix. The vector \(\hat\tau\) is referred to as the minimum mode, or the softest mode of the Hessian.

The direction \(\hat\tau\) is found iteratively using a finite difference between two replicas of the system, \(\mathbf{R}_{1,2} = \mathbf{R}_0 \pm \delta R \hat\tau\), around a known midpoint configuration \(\mathbf{R}_0\) starting with some orientation vector \(N\approx\hat\tau\). After the optimal orientation has been found, the dimer is moved in the direction opposite to the transformed gradient.
The saddle point search algorithm alternates between estimating the minimum mode by rotating the dimer and translating the center of the dimer until
the atomic forces at the midpoint tend to zero within user defined tolerance \cite{henkelmanDimerMethodFinding1999,goswamiBayesianHierarchicalModels2025a}.

Both the rotation and the translation require the use of some optimization algorithm.
For rotation, often the conjugate gradient with Polak-Ribiere \cite{goswamiBayesianHierarchicalModels2025a}
is used, and the limited memory Broyden-Fletcher-Goldfarb-Shanno algorithm, L-BFGS for the translation \cite{kastnerSuperlinearlyConvergingDimer2008,sheppardOptimizationMethodsFinding2008}.
The convergence characteristics are determined most strongly by the choice of the maximum
translation step and the initial direction of the dimer.

\begin{figure}[htbp]
\centering
\includegraphics[width=.9\linewidth]{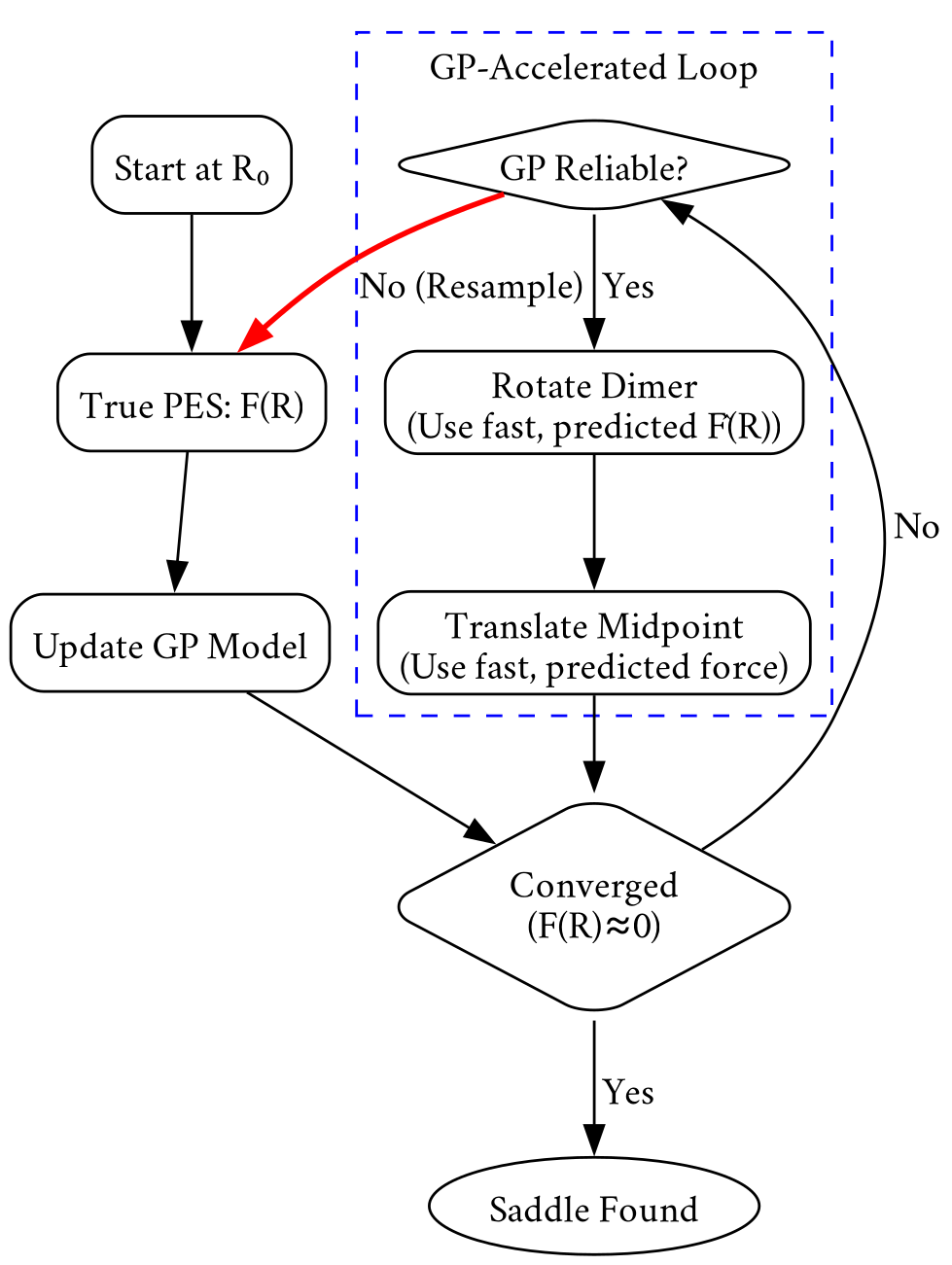}
\caption{\label{fig:gp_dimer_base}The previously developed GPDimer method for accelerating saddle point searches using Gaussian process regression. The GP generated surrogate surface is considered reliable if the guardrails of Eq. \ref{eq:mindistatm} and \ref{eq:max1dlog} are satisfied.}
\end{figure}

The Gaussian Process accelerated dimer, referred to here as GPDimer, involves constructing successive surrogate models of the PES on the fly.\cite{koistinenMinimumModeSaddle2020,goswamiEfficientImplementationGaussian2025a}
The GP approximation learns a function based on the previously computed points, typically electronic structure calculations for given positions of the atoms, and a set of optimized hyperparameters which maximize the likelihood of seeing the previous data:
\begin{equation}
V_{\text{GP}}(\mathbf{x}_{\text{new}} | M_{data}, \boldsymbol{\theta}_{\text{opt}}) \approx V(\mathbf{x}_{\text{new}})
\label{eq:gp_approx}
\end{equation}
where \(M_{data}\) form the observed pairs of \((V(x), -\nabla V(x))\) for \(x\) and
\(\mathbf{\theta}_{\text{opt}}\) are the hyperparameters of the kernel covariance
function. Fig. \ref{fig:gp_dimer_base} provides an overview of the algorithm.

The similarity between two configurations, \(\mathbf{x}\) and \(\mathbf{x}'\), is specified by the inverse-distance based squared exponential kernel \cite{koistinenMinimumModeSaddle2020,goswamiEfficientImplementationGaussian2025a}:
\begin{equation}
k(\mathbf{x}, \mathbf{x}') = \sigma_c^2 + \sigma_f^2 \exp\left(-\frac{1}{2} \sum_{i} \sum_{\substack{j > i}} \left(\frac{1/r_{ij}(\mathbf{x}) - 1/r_{ij}(\mathbf{x}')}{l_{\phi(i,j)}}\right)^2 \right)
\label{eq:idist_kernel}
\end{equation}
where \(r_{ij}\) defines the Euclidean distance between atoms \(i\) and \(j\), \(\sigma_f^2\) indicates the signal variance, along with the constant offset \(\sigma_c^2\), and the length scales \(l_{\phi(i,j)}\) for each atom pair type \(\phi(i,j)\). The \(1/r_{ij}\) term provides a strong physical prior, leveling out the sharp increase in repulsive force between pairs of atoms if the distance
between them becomes too short.

The electronic structure calculations at the most commonly used levels of methodology, such as density functional theory, provide the atomic forces with insignificant additional computational effort and thereby also the gradient of the energy. In order to make use of this information and more accurately reproduce the PES with the surrogate model surface, the gradients are included in the covariance matrix.
This requires extending the GP to a vector-valued framework. The joint distribution over predicted energy and atomic forces
involves a block covariance matrix:
\begin{equation}
\mathbf{K}_{\text{full}} = \begin{bmatrix} \mathbf{K}_{ff}(\mathbf{X}, \mathbf{X}) & \mathbf{K}_{f\nabla}(\mathbf{X}, \mathbf{X}) \\ \mathbf{K}_{\nabla f}(\mathbf{X}, \mathbf{X}) & \mathbf{K}_{\nabla\nabla}(\mathbf{X}, \mathbf{X}) \end{bmatrix}
\label{eq:full_covar}
\end{equation}
The additional blocks, such as \(\mathbf{K}_{f\nabla}\) and \(\mathbf{K}_{\nabla\nabla}\), represent the covariances between energy and forces, and between different force components, respectively, and derive from analytical derivatives of the base kernel \(k_{ff}\).
\cite{koistinenMinimumModeSaddle2020,goswamiEfficientImplementationGaussian2025a}

To ensure the GPDimer algorithm is stable and avoids unphysical regions of the PES,
\cite{koistinenMinimumModeSaddle2020,goswamiEfficientImplementationGaussian2025a}
two primary guardrails are employed.
First, a trust radius prevents the GP from extrapolating too far into unknown regions.
New configurations must fall within the defined radius relative to known
configurations in the ``1D max log'' distance metric, defined by:
\begin{equation}
D_{\text{1Dmaxlog}}(\mathbf{x}_1, \mathbf{x}_2) = \max_{i,j} \left| \log \frac{r_{ij}(\mathbf{x}_2)}{r_{ij}(\mathbf{x}_1)} \right|
\label{eq:max1dlog}
\end{equation}
In other words, the maximum absolute log-ratio of interatomic distances between two configurations determines the trust radius. This metric possesses the useful property of invariance to molecular rotations. However, its fundamental reliance on a fixed-index comparison between atoms creates a critical inconsistency with the physical symmetries of the system. This discrepancy motivates our later use of a permutation-invariant Optimal Transport measure.

Second, a constraint on the step-size prevents atoms from moving too close to each other as this would produce extreme atomic forces. The maximum allowed step length for any atom gets directly limited by its distance to its nearest neighbor, modulated by a \texttt{ratio\_at\_limit} parameter:
\begin{equation}
L_{\text{max}} = 0.5 \cdot (1.0 - \text{ratio}_{\text{limit}}) \cdot d_{\text{min}}
\label{eq:mindistatm}
\end{equation}
where \(d_{\text{min}}\) represents the minimum interatomic distance in the
current configuration. A \texttt{ratio\_at\_limit} value approaching \(1.0\) enforces a
highly cautious, stability-focused search with small steps, while a value
approaching \(0.0\) allows for more aggressive, larger steps to accelerate
convergence in well-behaved systems.
\subsection{Optimal Transport GP}
\label{sec:cmeth:otgp}
\begin{figure*}
\centering
\includegraphics[width=.9\linewidth]{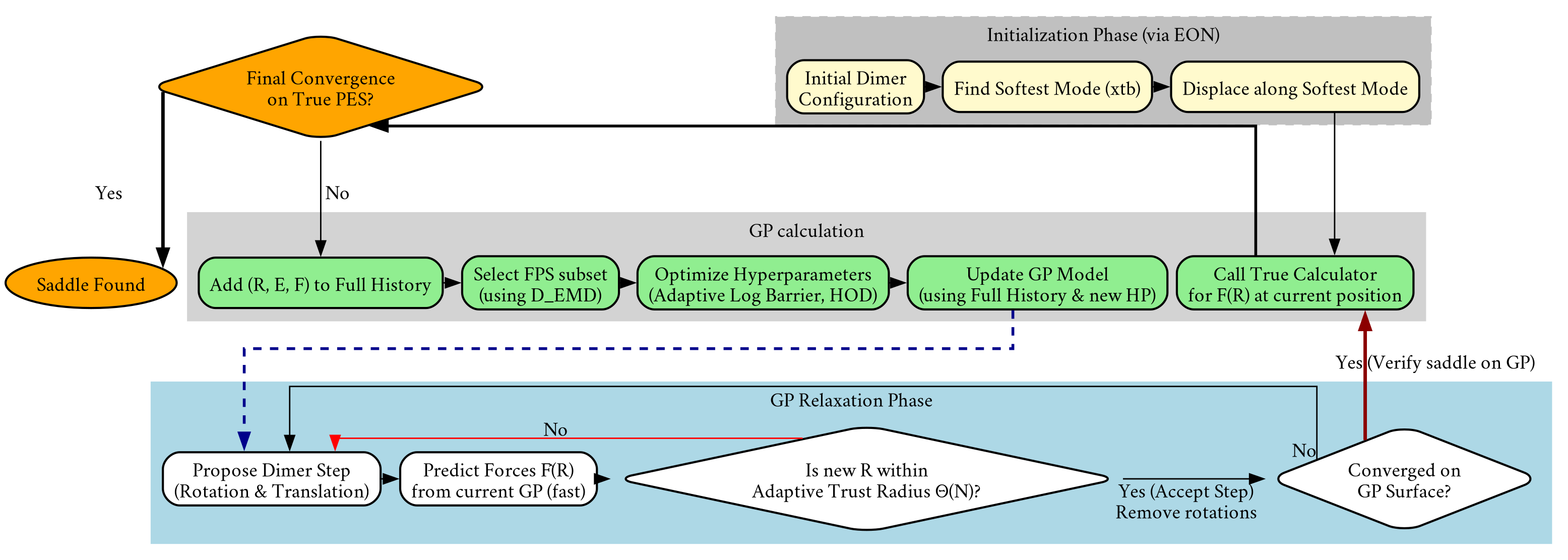}
\caption{\label{fig:otgp}Overview of the OT-GP method. Red arrows indicate rejected proposals (outside the trust radius); blue dashed arrows denote the transition from a freshly-updated GP model back to the relaxation loop; dark-red bold arrows highlight the final verification step on the true PES.  The flowchart therefore encapsulates the hierarchical control strategy of OT-GP: (i) cheap GP-driven exploration, (ii) data-driven trust-radius and variance regularisation, and (iii) intermittent high-fidelity validation that guarantees convergence to the first-order saddle point on the PES.}
\end{figure*}

The robustness and computational scaling of GP-accelerated searches link directly to the metric used to define dissimilarity between molecular configurations, \(D(\mathbf{x}_i, \mathbf{x}_j)\). Metrics that depend on fixed atom indices, such as the ``1D max log'' metric (Eq. \ref{eq:max1dlog}), lack invariance to the permutation of identical atoms. Even low-energy events, such as the rotations of a methyl group, may swap indices of two hydrogen atoms, resulting in a large, unphysical distance in this metric. To solve this, we ground our dissimilarity measure in Optimal Transport (OT) theory \cite{solomonOptimalTransportDiscrete2020,khanWhenOptimalTransport2022}. OT provides a deterministic prescription to ``lift'' a geometric distance between points into a distance between entire distributions of points. We treat each molecular configuration \(\mathbf{x}\) as a discrete distribution of element-typed (i.e., ``colored'') points in three-dimensional space. Conceptually, OT frames the distance calculation as a ``transport problem.'' If we consider a configuration \(\mathbf{x}_i\) to be piles of ``mass'' (the atoms) and configuration \(\mathbf{x}_j\) as a set of ``holes'' into which the mass must be moved, then the cost of moving one unit of mass from a point \(\mathbf{r}_a\) to a point \(\mathbf{r}_b\) is the Euclidean distance \(\|\mathbf{r}_a - \mathbf{r}_b\|\). The Earth Mover's Distance (EMD), the computational analogue of the Wasserstein-1 distance, finds the ``transport plan'' that minimizes the total work, i.e., the (mass \(\times\) distance) summed over all movements. By enforcing coloring, we have that ``carbon'' masses only move to ``carbon'' holes, respecting chemical identity. More importantly, the EMD inherently solves the permutation problem as the optimal transport plan corresponds to the optimal matching between the two sets of atoms, automatically assigning atom 1 in \(\mathbf{x}_i\) to its ``correct'' counterpart in \(\mathbf{x}_j\), regardless of its index. This yields a physically meaningful distance that correctly identifies symmetrically equivalent configurations as identical. We leverage this robust, permutation-invariant metric in our Farthest Point Sampling (FPS) to select a truly diverse set of configurations for hyperparameter optimization. Figure \ref{fig:otgp} summarizes the key modifications to the algorithm workflow in this improved GP acceleration of saddle point searches.

\subsubsection{Farthest Point Subsampling}
\label{sec:cmeth:fps}
The hyperparameter optimization required by a conventional GP requires a repeated inversion of a covariance matrix whose dimensions equal the total number of scalar degrees of freedom \(O((M_{data} \cdot N_{atoms})^3)\). To avoid this rapid scaling of computational effort, we instead construct a compact, maximally diverse training subset \(\mathcal{S}\subset\mathcal{X}\) (where \(\mathcal{X}\) denotes the full set of configurations for which the PES has been sampled) by means of Farthest Point Sampling (FPS) \cite{bartokMachineLearningUnifies2017}. At each iteration, the distance \(D(\mathbf{x}_i,\mathbf{x}_j)\) between every candidate configuration \(\mathbf{x}_i\in\mathcal{X}\setminus\mathcal{S}\) and all members of the current subset \(\mathcal{S}\) is evaluated. The next point is the one whose minimum distance to the subset attains the maximum.
\begin{equation}
\mathbf{x}_{\text{next}}
 = \arg\max_{\mathbf{x}_i\in\mathcal{X}\setminus\mathcal{S}}
   \Bigl[\min_{\mathbf{x}_j\in\mathcal{S}} D(\mathbf{x}_i,\mathbf{x}_j)\Bigr].
\label{eq:fps_samples}
\end{equation}
The application of Eq. \ref{eq:fps_samples} is repeated until a user-specified cardinality \(M_{\text{sub}}\) is reached. For guaranteed continuity of the dimer path, the two most recent configurations are always retained. For maximum efficiency, we avoid representations of atomic environments, and operate instead with Cartesian coordinates. However, the Euclidean distance measure for a molecule is a poor metric for distinguishability, since it lacks invariance to rotations or translations and possesses no concept of atomic types. We therefore introduce a new measure of distance for the FPS.
\subsubsection{Intensive Earth Mover's Distance}
\label{sec:cmeth:iemd}
The standard Earth mover's distance quantifies the
minimum work to transport one distribution of mass to another, resulting in an extensive property that scales with system size. We introduce an intensive formulation \(D(\mathbf{x}_i, \mathbf{x}_j)\) that respects chemical identity, permutation invariance, and size independence. This formulation allows the measure to be interpreted with chemical intuition and transferred across different molecules in a size-independent manner.

For a given atom type, \(t\)
a linear assignment problem is solved to match the atoms of that type between two configurations, yielding the per-type average displacement
\begin{equation}
\bar d_{t}
 = \frac{1}{N_{t}}
   \min_{\pi\in\Pi_{N_{t}}}
   \sum_{k=1}^{N_{t}}
   \bigl\|\mathbf{r}^{(1)}_{k,t}
         - \mathbf{r}^{(2)}_{\pi(k),t}\bigr\|.
\label{eq:emd_type_avgdisp}
\end{equation}
Here, \(N_{t}\) denotes the number of atoms of type \(t\) and \(\Pi_{N_{t}}\) the set of all permutations of the \(N_{t}\) indices. We then identify the largest per-type average displacement as the overall distance:
\begin{equation}
D(\mathbf{x}_i,\mathbf{x}_j)=\max_{t}\,\bar d_{t}(\mathbf{x}_i,\mathbf{x}_j).
\label{eq:emd_dist}
\end{equation}
Because each \(\bar d_{t}\) averages over the atoms of a particular element, it forms an intensive quantity that reflects the collective motion of a specific chemical group. Adding spectator atoms does not dilute the metric, which makes it an ideal measure for selecting a chemically diverse subset.

The hyperparameter vector, \(\boldsymbol{\theta}\), which includes length scales and signal variance, undergoes optimization by maximizing the marginal log-likelihood,
\begin{equation}
\log p(\mathbf{y} | \mathcal{S}, \boldsymbol{\theta}),
\label{eq:hypot_mll}
\end{equation}
on the FPS subset \(\mathcal{S}\). To ensure robust and efficient convergence, we implemented an iterative optimization scheme with key enhancements.
As with the earlier GPDimer, a Student's t-distribution prior is used, but for OTGPD we adapt the prior to apply stronger constraints as the chemical system's complexity increases, thereby preventing the model from overfitting. We set a floor of 28 for the prior's degrees of freedom to ensure a stable, near-Gaussian shape, and increase the value proportionally with the number of interaction types to enforce stronger regularization where needed.
\subsubsection{Computational Cost}
\label{sec:cmeth:ccost}
The FPS construction costs \(\mathcal{O}(M_{\text{sub}}M_{\text{data}})\) distance evaluations, each of which scales linearly with the number of atoms. For any fixed subset, hyper-parameter optimisation involves the inversion of a covariance matrix of size \((M_{\text{sub}}N_{\text{atoms}})\times(M_{\text{sub}}N_{\text{atoms}})\), i.e.
\begin{equation}
\mathcal{O}\bigl((M_{\text{sub}}N_{\text{atoms}})^{3}\bigr),
\label{eq:otgp_bigo}
\end{equation}
where \(M_{\text{sub}}\ll M_{\text{data}}\). An initial subset of 10 points is found to work well when used with adaptive growth of the history based on the hyperparameter oscillation and a hard user-defined upper limit of  \(M_{\text{sub}}=30\). To ensure that none of the computationally intensive PES calculations goes unused, the full history \(\mathcal{X}\) is used for the GP prediction of the energy on the surrogate surface.
\subsubsection{Hyperparameter Oscillation Detection}
\label{sec:cmeth:hod}
Re-optimization of the hyperparameters on a dynamically changing subset can lead to unstable estimates that oscillate between iterations. A hyperparameter oscillation detection (HOD) heuristic monitors these oscillations over a moving window of the last \(W\) optimization steps. We define an oscillation indicator, \(O(\theta_j, t)\), for each hyperparameter component \(\theta_j\) at optimization step \(t\):
\begin{equation}
O_{j}(t)=
\begin{cases}
1 & \text{if }\operatorname{sgn}\bigl[\Delta\theta_{j}(t-1)\bigr]\neq
      \operatorname{sgn}\bigl[\Delta\theta_{j}(t-2)\bigr],\\[4pt]
0 & \text{otherwise}.
\end{cases}
\label{eq:hod_cases}
\end{equation}
where \(\Delta\theta_j(t) = \theta_j(t) - \theta_j(t-1)\). We then compute the total oscillation fraction, \(f_{\text{osc}}\), over the window.
\begin{equation}
f_{\text{osc}} = \frac{1}{D(W-2)}\sum_{t=2}^{W-1}\sum_{j=1}^{D} O(\theta_j, t).
\label{eq:hod_cond}
\end{equation}
If \(f_{\text{osc}}\) exceeds a threshold,
\(p_{\text{osc}}\),
the optimization is flagged as being unstable.
In response, the size of the FPS subset, \(M_{sub}\), is incrementally increased and the optimization re-run, up to a maximum of three retries.
\subsubsection{Adaptive Barrier for Signal Variance}
\label{sec:scg_barrier}
The signal variance hyperparameter, \(\sigma_f^2\), governs the overall amplitude of the GP prior. When \(\sigma_f^2\) grows without bounds, the GP mean surface flattens, the predictive variance explodes, which in turn may drive the dimer into unphysical regions. To prevent this, we augment the marginal log-likelihood (MLL) objective function, with a logarithmic barrier:
\begin{equation}
\mathcal{L}_{\text{eff}}(\boldsymbol{\theta})
= \underbrace{\log p\bigl(\mathbf{y}\mid\mathcal{S},\boldsymbol{\theta}\bigr)}_{\text{MLL}}
  \;-\;
  \mu\,\log\bigl(\lambda_{\text{max}}-\log\sigma_{f}^{2}\bigr)
\label{eq:sigma_barrier}
\end{equation}
where \(\lambda_{\text{max}}>0\) fixes the absolute upper bound for \(\log\sigma_{f}^{2}\).
Here, we use \(\lambda_{\text{max}}=\log 2\) which corresponds to \(\sigma_{f}^{2}=2\) in the kernel units and \(\mu\ge 0\) controls the barrier strength.

The barrier strength grows linearly with the total number of collected data points,
\begin{equation}
\mu(N)=\mu_{0}+\alpha N,\qquad \mu(N)\le \mu_{\max},
\label{eq:sigma_linear}
\end{equation}
with \(\mu_{0}\) a small seed (e.g. \(10^{-4}\)), \(\alpha\) a modest growth factor (e.g. \(10^{-3}\)), and \(\mu_{\max}=0.5\) a user-defined ceiling. This schedule yields a weak barrier when the data is sparse to allow the GP to remain flexible and enforces an increasingly strict bound as the data set expands, thereby preventing runaway variance.

During optimization, the gradient of the barrier with respect to the optimization variable \(w_{0}=\log\sigma_{f}^{2}\),
\begin{equation}
\frac{\partial}{\partial w_{0}}\bigl[-\mu\log(C-w_{0})\bigr]
=
\frac{\mu}{C-w_{0}},
\label{eq:sigma_deriv}
\end{equation}
is added to the usual derivative of the MLL. The barrier therefore incurs negligible computational overhead while guaranteeing that the effective objective (Eq. \ref{eq:sigma_barrier}) remains finite and well-behaved throughout the hyper-parameter optimization.

The dimer calculations on the GP surface involve two additional enhancements, described below.
\subsubsection{Adaptive Trust Radius for Early Stopping}
\label{sec:tr_estop}
A new midpoint position of the dimer is accepted after a translation only if the atomic coordinates lie within a data-driven trust region. The distance metric is the intensive EMD defined in Eq. \ref{eq:emd_dist}.
Let
\begin{equation*}
d_{\text{EMD}}\bigl(\mathbf{x}_{\text{cand}},\mathbf{x}_{\text{nn}}\bigr)
\end{equation*}
be the EMD between the candidate configuration \(\mathbf{x}_{\text{cand}}\) and its nearest neighbour \(\mathbf{x}_{\text{nn}}\) in the current training set. We require
\begin{equation}
d_{\text{EMD}}\bigl(\mathbf{x}_{\text{cand}},\mathbf{x}_{\text{nn}}\bigr)\le
\Theta\bigl(N_{\text{data}},N_{\text{atoms}}\bigr),
\label{eq:nn_trust}
\end{equation}
where \(\Theta\) is an adaptive threshold that expands as the surrogate gathers information. The functional form follows an exponential saturation curve:
\begin{align}
\Theta_{\text{earned}}\bigl(N_{\text{data}}\bigr)&=
T_{\min }+\Delta T_{\text{explore}}\cdot\Bigl(1-e^{-k\,N_{\text{data}}}\Bigr), \\
k&=\frac{\ln 2}{N_{\text{half}}}.
\label{eq:nn_satcurv}
\end{align}
where \(T_{\min }\) denotes a minimal safe radius that prevents the dimer from taking trivially small steps, \(\Delta T_{\text{explore}}\) sets the maximal additional exploration distance that the algorithm may allow, and \(N_{\text{half}}\) corresponds to the number of reference points required for the threshold to reach 50\% of its maximum value, thereby controlling the rate of expansion.

A physically motivated ceiling prevents the trust radius from becoming unrealistically large for any system size, defined as
\begin{equation}
\Theta_{\text{phys}}(N_{\text{atoms}})
= \max\Bigl(\,a_{\text{floor}},\; \frac{a_{A}}{\sqrt{N_{\text{atoms}}}}\,\Bigr),
\label{eq:nn_trust_phys}
\end{equation}
with \(a_{\text{floor}}\) a user-defined lower bound and \(a_{A}\) a scaling constant that reflects typical inter-atomic separations.

The more restrictive of the two bounds forms the final criterion
\begin{equation}
\Theta(N_{\text{d,a}})
= \min\bigl(\Theta_{\text{earned}}(N_{\text{data}}),\,\Theta_{\text{phys}}(N_{\text{atoms}})\bigr).
\label{eq:nn_trust_exact}
\end{equation}
When a candidate violates the inequality in Eq. \ref{eq:nn_trust}, the algorithm rejects the step, performs an evaluation of the PES at the current midpoint, adds the new data point to the training set, and recomputes \(\Theta\). This targeted data acquisition actively improves the surrogate model where it proves to be unreliable, ensuring the step size adapts continuously to the knowledge encoded in the surrogate model.
\subsubsection{Rotation Removal}
\label{sec:rotrem}
Since the dimer method involves a projection of the atomic forces, Newton's third law is not necessarily obeyed even though the forces obtained from the PES do.
The system as a whole could, therefore, translate and rotate.
The GP model does not inherently enforce rotational or translational symmetry so net rotations and translations could influence its predictions. To prevent this this, our algorithm purifies each proposed translation step, \(\mathbf{s}\), by projecting out these
displacements. We calculate the pure internal step, \(\mathbf{s}_{\text{int}}\), by subtracting the component of the step that lies in the subspace of rigid-body motion:
\begin{equation}
\mathbf{s}_{\text{int}} = \mathbf{s} - \sum_{k=1}^{6} (\mathbf{s} \cdot \mathbf{u}_k) \mathbf{u}_k
\label{eq:rotrem}
\end{equation}
where \(\{\mathbf{u}_k\}\) represents an orthonormal basis that spans the external degrees of freedom. We generate this basis through a Gram-Schmidt orthonormalization of the infinitesimal rotation and translation basis vectors, as detailed in the SI (section S1.1).

This projection method provides an alternative to other approaches. For example, point-set registration techniques, such as those found in IRA \cite{gundeIRAShapeMatching2021}, or graph matching in d-SEAMS \cite{goswamiDSEAMSDeferredStructural2020}, can remove global rotations and translations but often incur significant computational costs. The GP kernel, Eq. \ref{eq:idist_kernel}, lacks permutation invariance, so enforcing it at the optimization step offers no real advantage. Quaternion-based methods \cite{melanderRemovingExternalDegrees2015} have
been found to be slower and less stable for the
optimization process \cite{goswamiBayesianHierarchicalModels2025a}.

The linear orthogonal projection requires only six inner products and a rank-6 update, incurring negligible overhead relative to the cost of evaluating the GP model, while ensuring that the dimer translation that proceeds to the next GP evaluation always uses \(\mathbf{s}_{\mathrm{int}}\).
\subsection{Computational Details}
\label{sec:cmeth:cdet}
The OT-GP algorithm is tested by accelerating the dimer minimum mode following search (OTGPD) on a benchmark of 238 organic molecules that contain one or two fragments, and range from 7 to 25 atoms derived from a set reported earlier \cite{hermesSellaOpenSourceAutomationFriendly2022,goswamiEfficientImplementationGaussian2025a}. The \texttt{eon} software \cite{chillEONSoftwareLong2014} is used to generate the
initial orientation of the dimer
and provides connection to
the electronic structure calculation, as in the previous
GP-Dimer implementation \cite{goswamiEfficientImplementationGaussian2025a}. All codes including the OTGP implementation and analysis scripts reside on Github. Materials Archive hosts generated data (see the SI).

The PES is obtained from Hartree-Fock (HF) calculations with a 3-21G basis set using the NWChem \cite{apraNWChemPresentFuture2020} software package interfaced through the \texttt{i-pi} \cite{kapilIPI20Universal2019} server-client optimizer interface. The calculations account for the spin state, unrestricted HF for doublets and restricted HF for singlets. For approximate and inexpensive calculations, such as the initial
orientation of the dimer
and Mayer-Wiberg bond-order analysis for fragments \cite{wibergApplicationPoplesantrysegalCNDO1968}, we use \texttt{ase} \cite{larsenAtomicSimulationEnvironment2017} with \texttt{xtb} and the \texttt{tblite} package for the \texttt{GFN2-xTB} \cite{bannwarthExtendedTightbindingQuantum2021,bannwarthGFN2xTBanAccurateBroadly2019} semi-empirical method.

Snakemake \cite{molderSustainableDataAnalysis2021} orchestrates runs on computer cluster
with Intel Xeon Platinum 8358 CPUs. The saddle point calculations are deemed converged when the root-mean-square atomic force falls below 0.01 eV/\r{A}. Runs terminate upon exceeding a 4-hour time limit, or if NWChem terminates with an error. The SI contains additional details.
\section{Results}
\label{sec:results}
\subsection{Benchmark Calculations}
\label{sec:res:bench}
\subsubsection{Reliability}
\label{sec:res:bench:success}
\begin{figure}[htbp]
\centering
\includegraphics[width=1.01\linewidth]{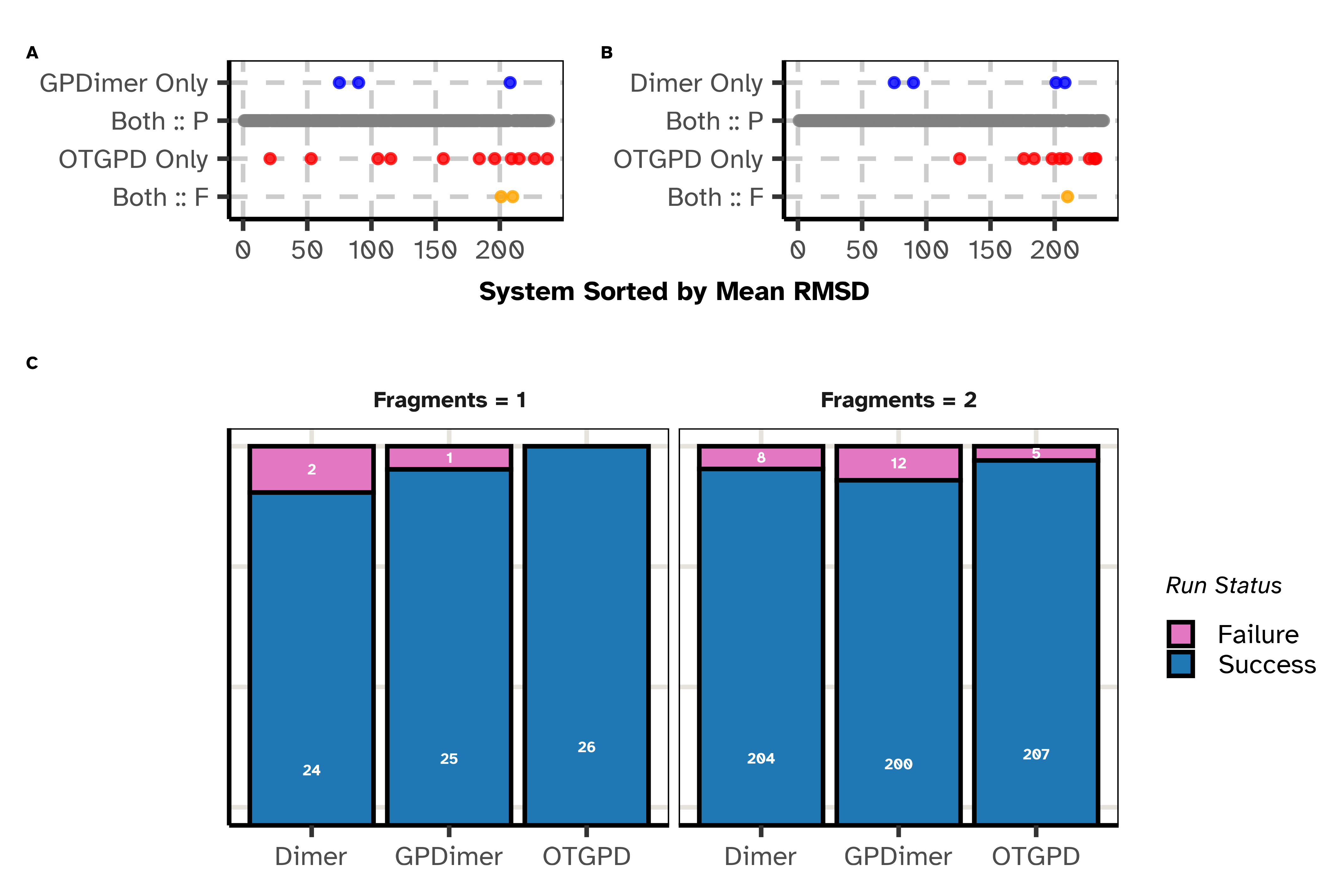}
\caption{\label{fig:otpgp_success}Comparison of the performance of the OTGPD saddle point search method against GPDimer and regular dimer without GP acceleration for 238 molecular systems. A calculation exceeding 4 hours or raising an error in the electronic structure calculation counts as a failure. (A, B) Success outcomes for each system, ordered along the horizontal axis by mean root-mean-square deviation (RMSD) between the saddle point and the initial atomic structure. Red dots denote systems where only OTGPD succeeded, blue dots where only the alternate method succeeded, and orange dots where both failed. (C) Bar chart summarizing success rate of each method for initial structures that represent a single fragment and those representing two fragments. The single-fragment cases represent typical saddle point searches, while the two-fragment cases test performance on more complex, often arbitrary dissociation pathways. The OTGPD method demonstrates a clear advantage. In addition to a mutual success rate exceeding 93\% in both comparisons, OTGPD uniquely finds the saddle point for an additional 11 systems (4.6\%) compared to GPDimer (1.3\%) and additional 9 systems (3.8\%) compared to Dimer (1.7\%).}
\end{figure}

The performance of the OT-GP method is compared against the previous GPDimer method and the original dimer method without GP acceleration on a benchmark set of 238 molecular systems, constructed by Hermes and coworkers.\cite{hermesSellaOpenSourceAutomationFriendly2022}
Figure \ref{fig:otpgp_success} summarizes the results of this comparison.

The comparison in Figure \ref{fig:otpgp_success} (A, B) reveals significant performance advantage of the OTGPD. While all three methods find a high degree of success (over 93\%), demonstrating the general utility of dimer-based approach, the critical differentiator appears in the cases where only one method succeeds. OTGPD uniquely finds the correct saddle point on 4.6\% of the systems compared to GPDimer, which in turn only succeeds alone on 1.3\% of the cases. This trend continues against the regular dimer method, where unique success rate of the OTGPD, 3.8\%, more than doubles that of the dimer, at 1.7\%. These statistics show that
OTGPD
succeeds on a significant number of
systems where the others fail.

An analysis of the failure modes in Figure \ref{fig:otpgp_success} (C) provides further insight. For the 26 single-fragment systems, which represent typical
starting configurations of the atoms,
OTGPD achieves a perfect success record with zero failures.
The systems where the bonding analysis of the initial atomic structure identifies two separate, non-bonded fragments are more
challenging and quite unusual as starting points for saddle point searches.
In most of these systems, the two fragments are far enough apart that the atomic forces are attractive and a minimization of the
energy would bring them closer to each other. This displacement typically represents the minimum mode. The saddle point search method, however, is designed to drive the system against the force acting along the minimum mode and therefore pushes the
fragments further apart, at least in the first few iterations. The path traced out by in the saddle point search eventually can turn around due to the minimization of the energy in orthogonal degrees of freedom, but this is indeed a highly unusual feature
of benchmark data for saddle point searches.\cite{hermesSellaOpenSourceAutomationFriendly2022}
We, nevertheless, choose to include these two-fragment systems in our tests, but exclude systems where the bonding analysis
detects three or more fragments.
OTGPD maintains the lowest failure rate of all three methods also for these two-fragment systems (5 failures, versus 12 for GPDimer and 8 for Dimer).

We further investigate the four systems (Fig. \ref{fig:otpgp_success} (A), blue dots) where the GPDimer runs succeed but the OTGPD runs fail. This discrepancy does not stem from a flaw in the OTGPD algorithm but from an artifact of the dimer initialization procedure (visualized in Fig. \ref{fig:otgp}). For OTGPD runs, we form the initial dimer configuration by displacing atoms along the softest mode found by an inexpensive semi-empirical (\texttt{xtb}, \texttt{GFN2-xTB}) calculation. In these specific cases, this displacement produced an unphysical starting geometry with high atomic forces or atoms in close contact. When any GP-accelerated method begins from such a high-energy, high-force baseline, the initial surrogate model learns high forces and energies corresponding to unphysical geometries are not anomalous, thus ``poisoning'' the run, leading to optimization failures. The GPDimer benchmark used a different initialization method which, for these systems, happened to produce a viable starting configuration. Robust dimer initialization routines are not a solved problem, nor the focus of this work. As detailed in the SI, when the OTGPD method starts from the same initial geometries used by the GPDimer, it converges successfully and more efficiently. For these four systems, the re-run OTGPD required a near-identical total number of HF evaluations (146 for OTGPD vs. 148 for GPDimer) while completing in less than half the average wall time (OTGPD, 51.3 minutes vs. GPDimer, 109.8 minutes), despite running with half the number of parallel processes (OTGPD 8 vs GPDimer 16).

\begin{figure}[htbp]
\centering
\includegraphics[width=.9\linewidth]{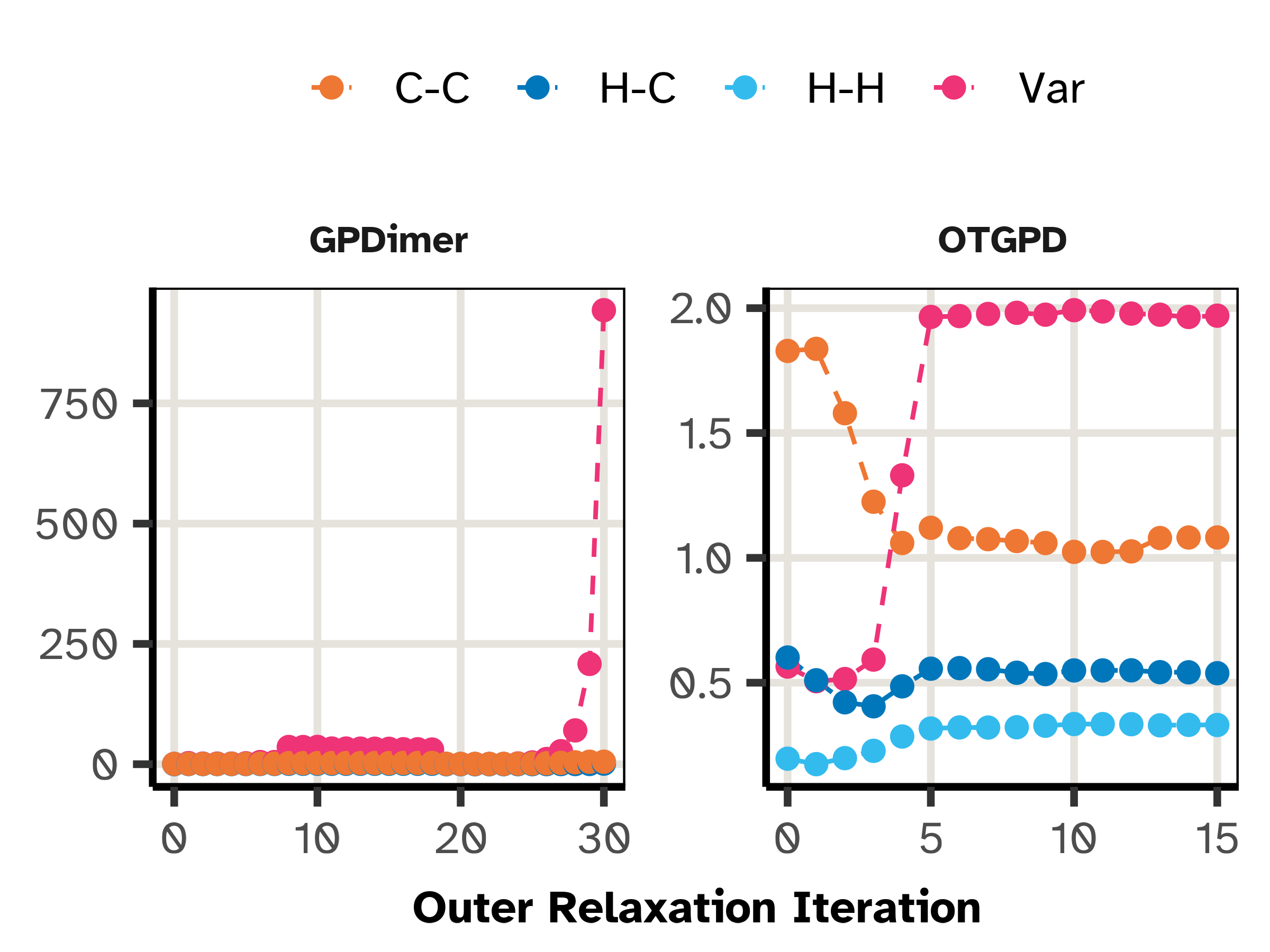}
\caption{\label{fig:var_explode}Evolution of the values for the hyperparameters in a representative failure case, system D110. (Left) The GPDimer search fails after 30 iterations as the signal variance (Var) explodes, leading to an error after 38 minutes. (Right) The OTGPD framework maintains a stable signal variance, which saturates at the adaptive barrier limit. This stability allows the search to converge successfully.}
\end{figure}

One of the calculations, the one for system D110,
is illustrated
in Figure \ref{fig:var_explode}. Here, the GPDimer search fails after its signal variance hyperparameter explodes, leading to a physically unrealistic configuration which crashes the electronic structure calculation by NWChem. In contrast, the OTGPD has stable values for the hyperparameters and locates the saddle point efficiently.

\begin{figure}[htbp]
\centering
\includegraphics[width=.9\linewidth]{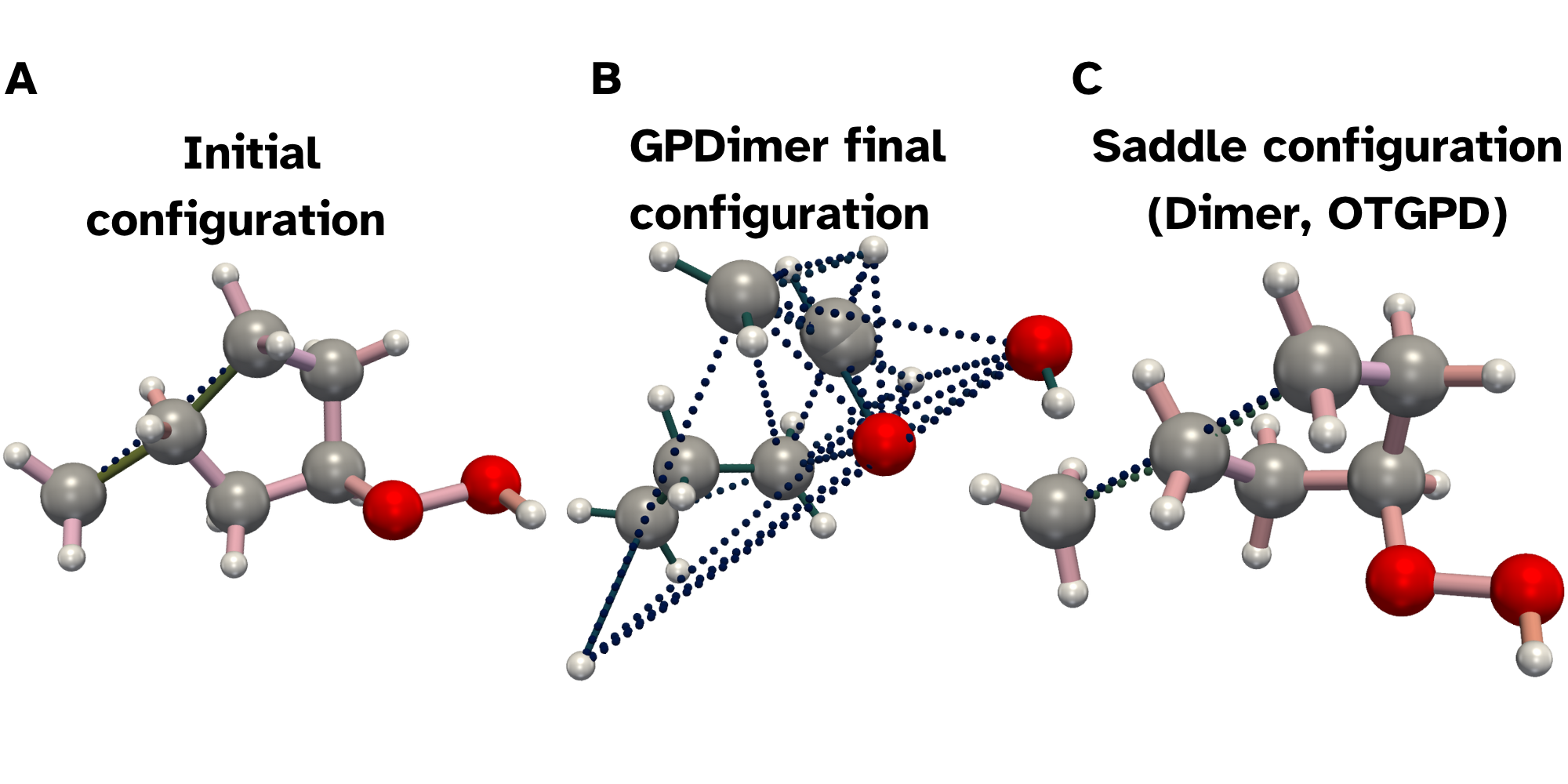}
\caption{\label{fig:d150_case_study}Comparison of saddle search outcomes for system D150, highlighting the stability of the OTGPD method. Bond coloring visualizes the Wiberg bond order, with dots used when the order is below 0.5, representing non-bonded interaction. (A) The initial geometry for the saddle point searches. (B) The GPDimer search fails and ends up fragmenting the molecule. (C) Both the standard dimer and the OTGPD method successfully converge to the same saddle point.}
\end{figure}

While the aggregate statistics demonstrate the OTGPD's superior reliability, individual cases provide a clearer physical intuition for the importance of its stability controls. System D150, a hydrogen abstraction reaction between a methyl and a 2-hydroperoxyethyl radical, provides an
example, illustrated in Figure \ref{fig:d150_case_study}.
Starting from
the initial geometry shown in Figure \ref{fig:d150_case_study} (A), the GPDimer search fails catastrophically, resulting in a 
fragmentation of the molecule into nine components as shown in Figure \ref{fig:d150_case_study} (B) with atoms colliding into each other. This failure mode exemplifies a surrogate model collapse, likely driven by an unconstrained variance explosion. In 
contrast, both the standard dimer calculation, illustrated in Figure \ref{fig:d150_case_study} (C) and the OTGPD correctly navigate the potential energy surface and converge on the same saddle point. This case study underscores that the innovations in OTGPD provide more than a speed increase; they introduce essential guardrails that ensure the reliability of the search, preventing the severe failures that characterized the previous generation of the GPR-accelerated saddle point search method.

\subsubsection{Saddle Consistency}
\label{sec:saddle_consistency}
We next assess the quality of the saddle points located by each method. A direct comparison of barrier heights proves
to be difficult
particularly for systems that begin as two fragments since the search path is indirect. For these cases, the long and curved search path can result in convergence to different saddle points depending on minor variations in the trajectory, making a one-to-one comparison of single runs inconclusive.

Our analysis, therefore, focuses on algorithmic consistency across the full set of mutually successful runs. When compared against the baseline dimer method, the OTGPD method demonstrates high fidelity, locating transition states of effectively equal energy in 89.7\% of cases. In the few instances with differing outcomes, the energy differences remain small (< 0.05 eV) and show no systematic bias toward either method. The comparison with GPDimer reveals a similarly high degree of agreement, with both methods finding
saddle points with the same energy in over 75\% of the systems.

These results show that significant gains in speed and reliability do not come at the expense of accuracy, as it consistently reproduces the results of the baseline dimer method.
\subsubsection{Performance and Computational Cost}
\label{sec:perf}
Having established the superior reliability of the OTGPD method, we now examine its computational efficiency. An ideal algorithm must not only work well but also work fast, minimizing both the total time-to-solution and the number of expensive electronic structure calculations.

\begin{figure*}
\centering
\includegraphics[width=.9\linewidth]{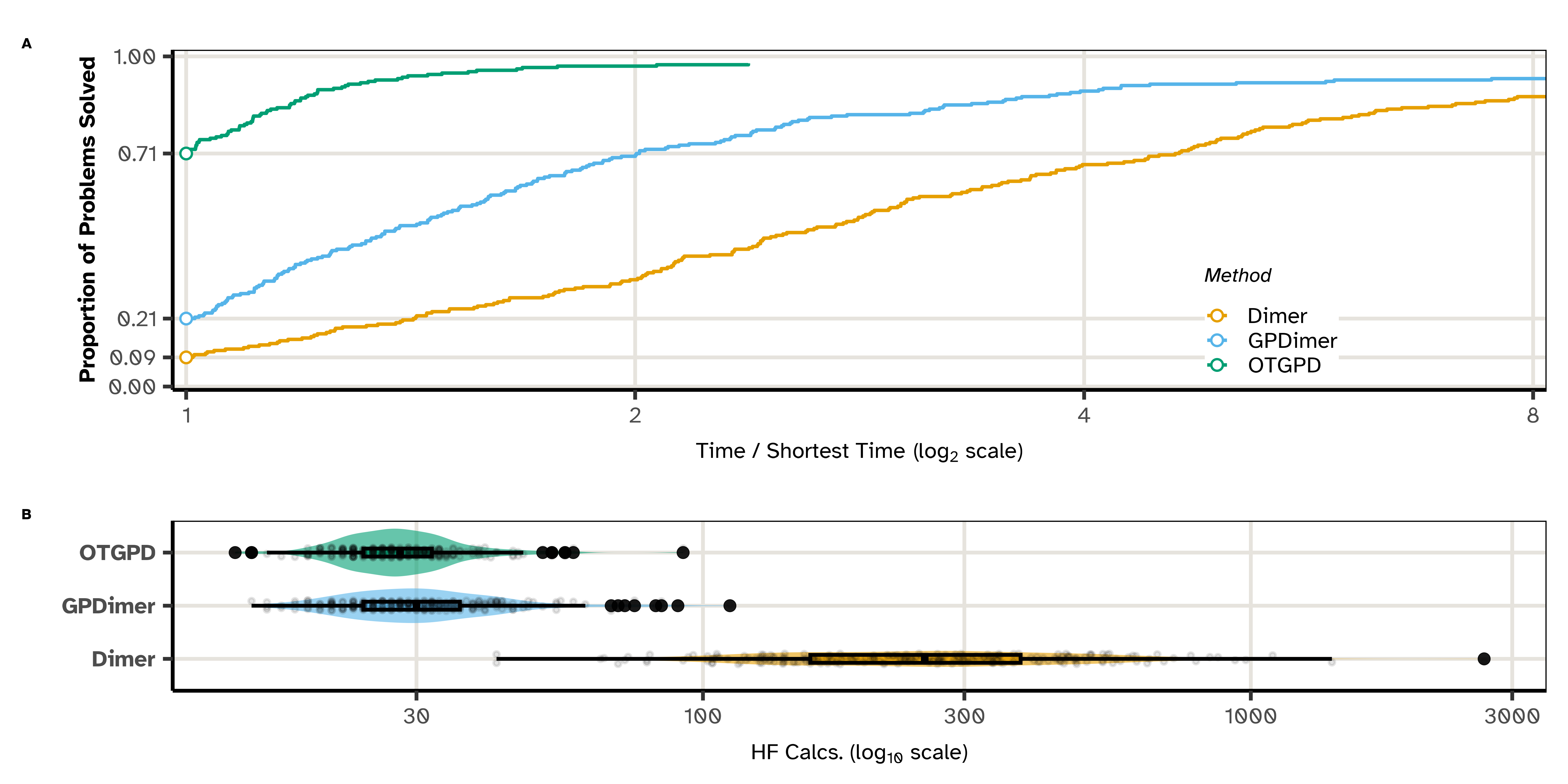}
\caption{\label{fig:otpgp_perf}Comparison of the computational efficiency of the OTGPD, GPDimer, and standard dimer methods. (A) A performance profile graph comparing the wall-clock time needed to reach the saddle point. The vertical axis shows the proportion of problems solved within a performance ratio \(r\) of the best time recorded for that system. A curve that is higher and further to the left indicates superior performance. The OTGPD method (teal) solves 70.6\% of problems with the best time (\(r\)=1), reflected in its median performance ratio of 1.00. This significantly outperforms both GPDimer (win rate 20.6\%, median \(r\)=1.39) and the standard dimer method (win rate 8.8\%, median \(r\)=2.65). (B) Violin and box plots showing the distribution of the number of HF calculations for the successful runs of each method (populations detailed in Figure \ref{fig:otpgp_success}). Over these valid systems, both GP-accelerated methods reduce the median number of HF calculations by an order of magnitude from 254 for the standard dimer method to 30 for GPDimer and 28 for OTGPD. These gains in data efficiency directly lead to the superior wall-time performance seen in (A), with OTGPD’s mean time to solution (12.6 min) reducing the wall time by nearly a half compared to the standard Dimer (23.7 min) and to less than a half compared to GPDimer (28.3 min).}
\end{figure*}

Figure \ref{fig:otpgp_perf} (A) presents a performance profile \cite{dolanBenchmarkingOptimizationSoftware2002}, a visualization for benchmarking solver efficiency which measures the proportion of best in class performance across systems. The profile for OTGPD dominates the other two methods at every performance ratio,
\(r\).
OTGPD solves over 70\% of the problems as the fastest method (\(r=1\)) and reaches its maximum success rate within a factor of two of the best time \((r\approx 1.5)\). In stark contrast, GPDimer requires four times the best time (\(r=4\)) to solve 90\% of its successful cases, and the standard Dimer lags even further behind.

While wall time reflects overall efficiency, the number of electronic structure calculations often represents the core computational bottleneck. In the calculations presented here a particularly fast electronic structure method is used, HF. In many cases the computational effort of each PES evaluation is much more involved and the overhead of the GP thereby smaller in comparison.
Figure \ref{fig:otpgp_perf} (B) shows the
the number of HF calculations for all successful runs. As previously demonstrated \cite{goswamiEfficientImplementationGaussian2025a}, GP-acceleration provides an order-of-magnitude reduction in this cost compared to the standard dimer method. The standard dimer method exhibits a wide distribution centered around 300 HF calculations. Both GPDimer and OTGPD shift this distribution down to a median of approximately 30 calculations. Notably, while the median number of HF calculations for OTGPD and GPDimer appear similar, the distributions and mean wall times reported in the caption reveal a crucial difference. A subset of long-running calculations skews the GPDimer performance, an issue that the algorithmic enhancements of the OTGPD resolves, resulting in a tighter performance distribution and a mean time-to-solution less than half that of GPDimer.

\begin{figure*}
\centering
\includegraphics[width=.9\linewidth]{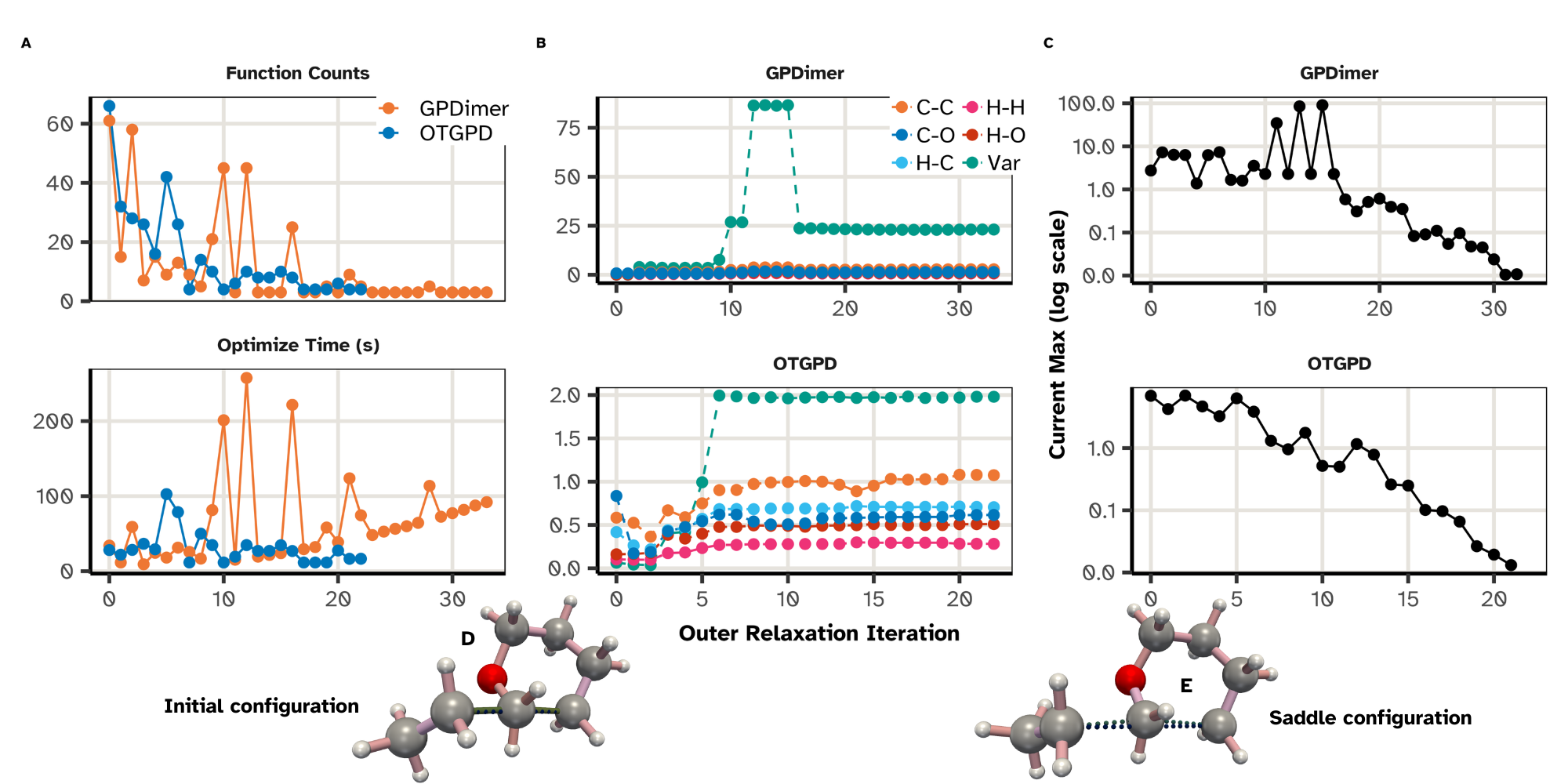}
\caption{\label{fig:d136}A detailed performance trace for a representative system, D136, for which OTGPD converges in 19.9 min versus 45.8 min for GPDimer. (A) The per-iteration cost of hyperparameter optimization, showing the number of electronic structure calculations (top) and wall time (bottom). The GPDimer trace (orange) exhibits large, erratic spikes that are absent for OTGPD (blue). (B) The evolution of the key hyperparameters. The GPDimer method shows a transient instability where the signal variance (teal) to a large value, coinciding with the cost spikes seen in (A). The adaptive barrier in OTGPD, however, maintains hyperparameter stability. (C) The convergence of the maximum force component on the true potential energy surface, showed on a log scale. The hyperparameter instability in the GPDimer calculation disrupts the smooth convergence of the geometry search. (D, E) The initial and saddle point configurations for the hydrogen abstraction reaction. For visual clarity, a Wiberg Bond Order cutoff of 0.5 highlights the key bond-breaking and bond-forming events. Ultimately, the combination of model stability and smooth convergence for the OTGPD allows it to find the saddle point using only 28 HF calculations, compared to the 39 required by the less stable GPDimer run.}
\end{figure*}

To understand the source of this performance difference, we examine the detailed optimization trace of a representative system, D136, illustrated in Figure \ref{fig:d136}. The GPDimer trace reveals large spikes in the per-iteration hyperparameter optimization time, which coincide with a transient, explosive increase in the signal variance and a temporary stall in force convergence. In stark contrast, the OTGPD trace shows stable optimization times and smooth, monotonic force convergence. This difference dramatically impacts the final performance: the OTGPD locates the saddle point in 19.9 minutes using 28 HF calculations, while the GPDimer needs 45.8 minutes and 39 HF calculations.

Crucially, this comparison remains valid despite the different initial orientation of the dimer. Both algorithms first perform a rotation on the PES, a step that equalizes the starting conditions by aligning the search with the physically correct softest mode. This procedure yields nearly identical starting geometries and therefore, similar GP models for the main search phase. The minor difference (\(\approx 2\) calls) during this initial rotation vanishes in the context of the overall improvement of 11 calls for the OTGPD. The OTGPD takes 20 minutes, compared to the 45 minutes for the GPDimer. Therefore, the primary performance gains stem directly from the superior stability and efficiency of OTGPD’s subsequent optimization steps. See also a detailed comparison in the SI.

Finally, a hierarchical Bayesian analysis \cite{goswamiBayesianHierarchicalModels2025a} provides a rigorous confirmation of these findings (details in the SI). The model shows that OTGPD reduces the median time-to-solution to less than a half compared to the dimer method (9.0 min vs 20.5 min) and runs \(\approx\)30\% faster than the GPDimer (9.0 min vs 12.8 min), with a greater than 99\% probability that OTGPD outperforms the other two methods.
\section{Discussion}
\label{sec:discussion}
The efficiency of the OTGPD method arises from a hierarchical system of data-driven controls that stabilize both the surrogate model and the geometry search it guides. By analyzing the interplay between our methods for hyperparameter optimization and the mechanisms governing the dimer translation, we can develop a physical intuition for the algorithm's robustness and speed.

We first consider the ideal scenario for a GPR-accelerated search. An ideal surrogate would learn the true potential-energy surface from sparse data; the dimer based saddle point search would then converge rapidly with stable hyperparameters, and the GP would act as a perfect accelerator with no wall-time overhead. In practice, this ideal fails for two fundamental reasons.

First, the surrogate operates under
data scarcity. A typical saddle search provides the model with only a few dozen training points, many of which represent unphysical arrangements of the atoms. This provides insufficient information to learn the true, high-dimensional energy landscape, and is in sharp contrast to machine-learned potential functions that rely on thousands of data points \cite{khanKernelBasedQuantum2023} using specialized features \cite{musilPhysicsInspiredStructuralRepresentations2021}. Second, the GP's marginal-likelihood objective possesses an intrinsically shallow ridge in the signal-variance direction. This allows the optimizer to increase \(\sigma_{f}^{2}\) arbitrarily to accommodate strained outlier configurations. This uncontrolled variance growth leads to a flattening of the mean surface and an explosion of the predictive variance, which in turn guides the saddle point search algorithm to propose unphysical steps that can lead to failures in the electronic structure calculation.
\subsection{Improved Wall-time Efficiency}
\label{sec:disc:eff}
For all GP regression models, hyperparameter optimization dominates the wall time because the covariance matrix grows with every new configuration \cite{mackayInformationTheoryInference2019,rasmussenGaussianProcessesMachine2004}. The inversion cost scales as
\[
\mathcal{O}\bigl((M_{\text{data}}3N_{\text{atoms}}+1)^{3}\bigr).
\]
The optimization step rapidly eclipses the cost of the underlying calculation, even when the optimization converges in a few steps since the hyperparameters do not change much once enough data accumulates to allow the dimer to converge on the GP surface. Figure \ref{fig:d136} (A) shows that after 20 steps, even when the hyperparameters do not change, single evaluations of the hyperparameter objective for the GPDimer (orange) grows steadily in wall time. By introducing the FPS/EMD pipeline described in sections \ref{sec:cmeth:fps} and \ref{sec:cmeth:iemd} we perform the hyperparameter optimization on a fixed-size, chemically diverse subset of data which scales as
\[
\mathcal{O}\bigl((M_{\text{sub}}3N_{\text{atoms}}+1)^{3}\bigr).
\]
Practically, this means a more than two-fold reduction in mean wall time (12.6 minutes versus 28.3 min for the previous GP-Dimer) while preserving the order-of-magnitude savings in electronic structure calculations. The accuracy and stability are improved because no sub-sampling of the data through inducing points \cite{deringerGaussianProcessRegression2021} or nesting approximations to the GP  \cite{denzelGaussianProcessRegression2019,gramacySurrogatesGaussianProcess2020} takes place.

The most common approach to handling GPR scaling involves sparse (inducing point) methods \cite{rasmussenGaussianProcessesMachine2006,gramacySurrogatesGaussianProcess2020}. These methods work well for large, static datasets, however, our ``on-the-fly'' design lies within a fundamentally different, dynamic regime (Fig. \ref{fig:otgp}). The dynamic re-optimization reduces the advantages of standard sparse GPR which involves an additional optimization at each training step to construct the inducing point locations. We estimated this would prove prohibitive and likely negate any savings, given our median sample size of approximately 30 PES evaluations.

The FPS method confines the expensive hyperparameter optimization to a small subset (\(M_{\text{sub}}\)) while still using the full data history (\(\mathcal{X}\)) for the predictive step. This ``data selection'' approach maintains a link to the physical system as the FPS subset consists of observed configurations that show maximal diversity in EMD metric. These points ``bracket'' the relevant region of the PES, an interpretation which is harder in a sparse-GP context, where the inducing points represent optimized mathematical constructs. We anticipate sparse methods may prove beneficial for much larger problems requiring hundreds of evaluations, but not for the class of problems considered here.
\subsection{Interpretation of the Hyperparameters}
\label{sec:disc:hyp}
A key insight into the GPDimer method concerns the physical meaning, or lack thereof, of its hyperparameters. One might believe that the length scales, \(l_{\phi(i,j)}\) in Eq. \ref{eq:idist_kernel}, of a Gaussian Process \cite{deringerGaussianProcessRegression2021} correlate to physical quantities such as covalent bond radii \cite{kaappaGlobalOptimizationAtomic2021,garijodelrioMachineLearningBond2020}. This view, however, lacks a first-principles justification. We define the kernel's length scales per atom-pair type, for example, one for C-C, C-H, etc.
not per individual covalent bond. Because we optimize these parameters over all instances of a pair type in the training subset \(\mathcal{S}\), they typically do not converge to specific equilibrium bond lengths. Instead, they represent a global, averaged ``stiffness'' for each interaction type, which represents the local PES region explored at any given moment. The hyperparameter vector \(\boldsymbol{\theta}\) thus contains these length scales and the signal variance, \(\sigma_f^2\) shown in Eq. \ref{eq:idist_kernel}. The GP kernel uses these few parameters to learn complex, higher-order, many-body effects implicitly.

Recall that our fundamental goal in a saddle point search involves finding a stationary point, \(x^*\), on the true, but slow to evaluate, PES. This corresponds to locating a configuration where the atomic forces, used to drive the iterations in Eq. \ref{eq:dimer_base}, vanish. In terms of the many-body expansion (MBE) \cite{stoneTheoryIntermolecularForces2013,musilPhysicsInspiredStructuralRepresentations2021}:
\begin{equation}
V(x) = \sum_i V_1(i) + \sum_{i<j} V_2(i, j) + \sum_{i<j<k} V_3(i, j, k) + \dots + V_N(1, \dots, N)
\label{eq:mbe}
\end{equation}
where \(V_1\) represents the energy of isolated atoms, \(V_2\) captures all pairwise interactions, \(V_3\) the three-body effects, and so on, and higher order terms are non-negligible \cite{pozdnyakovIncompletenessAtomicStructure2020}.
The computational cost of sampling \(V(x)\) motivates the substitution of a GP surrogate model, \(f(x; \boldsymbol{\theta})\), for \(V(x)\), as shown in Eq. \ref{eq:gp_approx}. We then transfer our physical objective to this surrogate, seeking a solution to:
\begin{equation}
\nabla f(x; \boldsymbol{\theta}) \approx 0
\label{eq:surrogate_objective}
\end{equation}
Note that the hyperparameter objective function of a GP model, however, describes something fundamentally different than Eq. (\ref{eq:mbe}). A GP defines a probability distribution over functions. After observing a set of \(N\) data points, \(\mathcal{S}\), our surrogate surface, \(f(x; \boldsymbol{\theta})\), corresponds to the mean of a posterior multivariate normal (MVN) distribution, conditioned on that data.

A critical disconnect emerges in how we select the hyperparameters, \(\boldsymbol{\theta}\). \(\boldsymbol{\theta}\) cannot minimize the true error between the surrogate and PES, which would require extensive sampling the PES and defeat the purpose of the surrogate approach:
\begin{equation}
\boldsymbol{\theta}_{\text{ideal}} = \arg\min_{\boldsymbol{\theta}} \int |f(x; \boldsymbol{\theta}) - V(x)|^2 dx \quad \text{(Inaccessible)}
\label{eq:ideal_theta}
\end{equation}
Instead, we select \(\boldsymbol{\theta}\) by maximizing the marginal log-likelihood (MLL), a purely statistical quantity derived from our sparse training subset \(\mathcal{S}\), selected here via Farthest Point Sampling. This makes our actual hyperparameter objective:
\begin{equation}
\boldsymbol{\theta}^* = \arg\max_{\boldsymbol{\theta}} \log p(\mathbf{y} | \mathcal{S}, \boldsymbol{\theta})
\label{eq:mll_theta}
\end{equation}
which does not correspond directly to the formal physical series in Eq. (\ref{eq:mbe}) with a single function realized from an MVN distribution. The process of finding optimal hyperparameters by maximizing the marginal log-likelihood, \(p(\mathbf{y} | \mathcal{S}, \boldsymbol{\theta})\), which measures the model's consistency, under the assumption the underlying data follows a multi-variate normal distribution, conditioned on the sparse data it has seen. No other direct information about the Eq. (\ref{eq:mbe}) enters the algorithm.

Thus, the standard approach implicitly relies on a chain of unjustified assumptions:
\begin{align}
\underbrace{\arg\max_{\boldsymbol{\theta}} \log p(\mathbf{y} | \mathcal{S}, \boldsymbol{\theta})}_{\text{We optimize this...}} &\implies \boldsymbol{\theta}^* \\
&\implies \underbrace{f(x; \boldsymbol{\theta}^*) \approx V(x)}_{\text{...assuming this...}} \\
&\implies \underbrace{\nabla f(x^*) \approx 0 \text{ gives } \nabla V(x^*) \approx 0}_{\text{...to achieve this.}}
\label{eq:assumption_chain}
\end{align}
We optimize a statistical property and hope it reproduces a physical reality it was never designed to model. Our work's adaptive controls act as a corrective. They impose functional constraints on the function drawn from the GP, serving as a proxy for the unknown physical constraints that the statistical objective function lacks.

This perspective also clarifies the role of the signal variance, \(\sigma_f^2\), which governs the overall flexibility of the surrogate model. Our analysis reveals that unstable or failed searches often correlate with an unphysically large signal variance. Such a large variance allows the model too much freedom to oscillate between energetically similar but geometrically distinct configurations. This physical insight directly motivates the adaptive barrier method described in Section \ref{sec:scg_barrier}, which constrains the variance to a physically reasonable range, thereby greatly enhancing the stability and robustness of the search.
\subsection{Hyperparameter Guardrails}
\label{sec:disc:hypguard}
Having established that the hyperparameters lack a direct physical interpretation, we now consider the data-driven guardrails essential for the OTGPD method. The marginal-likelihood landscape for a GP surrogate often presents a shallow ridge in the signal-variance direction. Without regularization, the optimizer can increase \(\sigma_{f}^{2}\) arbitrarily as new, strained
configurations of the atoms enter the training set.

In GPDimer calculations, this is manifested as a rapid, uncontrolled growth of \(\sigma_{f}^{2}\), illustrated in Figure \ref{fig:var_explode} (A). As the signal variance increases, the model becomes pathologically flexible. It ceases to act as a physical model and instead behaves as a pure mathematical interpolator between sparse and increasingly high energy reference data. This compliance means the surrogate sees no penalty for proposing physically impossible configurations in efforts to honor the fit. This guides the saddle point search to unphysical arrangements of the atoms, such as overlapping atoms, causing the electronic structure calculation to eventually fail.

This specific failure mode requires a solution distinct from conventional L1 or L2 regularization \cite{jamesIntroductionStatisticalLearning2013}. Such methods apply a soft penalty to the objective to combat general overfitting, but a strong MLL gradient can overcome this penalty, failing to prevent the variance explosion seen in Fig. \ref{fig:var_explode}. We therefore draw inspiration from interior point methods \cite{potraInteriorpointMethods2000} used in constrained optimization.

Embedding the logarithmic barrier of Eq. (\ref{eq:sigma_barrier}) directly into the objective function serves as a ``hard wall'' to enforce this constraint. Unlike a soft penalty, the barrier creates a diverging gradient (Eq. \ref{eq:sigma_deriv}) as the optimizer approaches the bound \(\lambda_{\max}\), which guarantees that the optimizer cannot enter the unstable region. Such constraints on \(\sigma_{f}^{2}\) are interpretable, with a physically meaningful upper bound rather than a more abstract L1/L2 penalty strength. We find this gradient-based barrier necessary as a simple hard bound proved insufficient, with the optimizer ``sticking'' to the boundary.

Because the barrier strength \(\mu\) scales with the data count, Eq. (\ref{eq:sigma_linear}\}), the surrogate remains flexible in the early steps. This allows the optimisation to explore a wide variance range and progressively tightens as the data set matures, exactly when the model should have settled on a physically reasonable amplitude. This adaptive behaviour eliminates the pathological growth of the variance while preserving the surrogate model's ability to capture the curvature of the true PES, as seen in Figure \ref{fig:var_explode} (B).

The HOD mechanism, Eqs. (\ref{eq:hod_cases}) and (\ref{eq:hod_cond}, monitors the stability of all hyper-parameters, not only \(\sigma_{f}^{2}\). Whenever frequent sign changes appear in the update history, the FPS subset enlarges, and if there are several high energy points, not all are taken, which improves the conditioning of the covariance matrix and yields a smoother marginal-likelihood surface. Empirically, the combination of barrier and HOD reduces the number of failed saddle searches from roughly twelve percent (baseline GPDimer) to two percent for the OTGPD method, while the median wall-time drops from twenty-eight to twelve minutes across the 238-molecule benchmark.
\subsection{Step-size Control}
\label{sec:disc:stepsize}
While the translational step size is an important parameter in any saddle point search, its strategic importance on a surrogate surface gives way to the model's own adaptive, data-driven guardrails. In a conventional search, where each step requires an expensive calculation, the step size provides the user's primary, manually tuned lever for balancing progress and stability. Choosing an excessively large step size can destabilize the search, causing the system to overshoot the nearest saddle point and explore unphysical, high-energy regions of the energy surface, often leading to convergence failure. On surrogate surfaces, new samples have negligible computational cost. Thus the step size can take a small, conservative value without penalty. The primary driver for controlling the search therefore shifts from the step size to the adaptive parameters of the surrogate model itself.

The stability of the search is critically dependent on the distance metric used to define the trust radius. The previous GPDimer approach used the ``1D max log'' metric given by Eq. (\ref{eq:max1dlog}), which is not invariant to the permutation of identical atoms. This flaw means that a low-energy symmetric event, such as the rotation of a methyl group or proton transfer, can be misinterpreted as a large geometric step. This may incorrectly halt the search or add redundant symmetric configurations to the training set, which leads to ill-conditioned kernel matrices and numerical instability.
By defining the trust radius with our intensive Earth Mover's Distance (\(D_{\text{EMD}}\) from Eq. (\ref{eq:emd_dist})), we impose a physically meaningful and transferable criterion for a physically motivated, automated step-size controller. The metric's intuition, which corresponds to the average displacement of atom groups ensures that accepted steps correspond to sensible chemical transformations, not 
unphysical leaps. A proposed configuration that falls outside the trust region represents the direct analogue of taking an excessively large step in a traditional saddle point search.

The adaptive threshold \(\Theta(N_{\text{data}})\) embodied in Eq. (\ref{eq:nn_trust}) embodies an ``earned-exploration'' principle. At the start of a search with only a few reference points, \(\Theta\) remains close to the minimal safe radius \(T_{\min }\) so the algorithm thus takes conservative steps. As the surrogate accrues more data, \(\Theta\) grows smoothly toward \(T_{\min }+\Delta T_{\text{explore}}\), permitting larger moves
when the model has become sufficiently accurate. The exponential form ensures rapid initial growth that tapers off, preventing runaway step sizes in later stages.

The physical ceiling \(\Theta_{\text{phys}}\), Eq. (\ref{eq:nn_trust_phys}) safeguards against pathological expansions of the trust radius on systems with many atoms or on early iterations where the earned term could otherwise dominate. By scaling as
\(a_{A}/\sqrt{N_{\text{atoms}}}\) the bound reflects the typical size of atomic displacements and ensures that the trust region remains chemically reasonable irrespective of system size. Taking the minimum of the earned and physical limits, Eq. (\ref{eq:nn_trust_exact}) therefore guarantees that every accepted step respects both data-driven confidence and universal geometric constraints.

By rejecting overly ambitious proposals early, the algorithm avoids failed electronic structure evaluations, which can be costly, and limits the number of additional reference points required to bring the surrogate back into a trustworthy regime. Consequently the median number of electronic structure calculations remains low while halving the average time-to-solution between the GPDimer and OTGPD.
\subsection{Removal of Rotation}
\label{sec:disc:rotrem}
Any surrogate model that operates in Cartesian coordinates must distinguish genuine internal deformations from translation and rotation
of the system as a whole. If the GP model receives a step containing a net translation or rotation of the system, the kernel evaluates distances between atoms that have undergone a uniform shift. Because the inverse-distance kernel treats every Cartesian component independently, such a shift introduces artificial changes in the pairwise distances. These artifacts in turn may lead to spurious force predictions and an artificial increase of the predictive variance.

By projecting each step onto the internal subspace eliminates this source of error. By construction, this procedure guarantees that the updated atom coordinates share the same center of mass and orientation as the previous configuration. The pairwise distances that enter the kernel therefore remain unchanged apart from the true internal deformation. The GP consequently observes a smooth evolution of the input space, which stabilizes both the mean prediction and the variance.
While a general best practice for any GP-accelerated search, implementing this purification step contributes to the high success rate of our OTGPD method. This enhanced stability manifests in the marked reduction of failed runs (from twelve percent to less than two percent on our benchmark) and in the tighter distribution of wall-time values.
\section{Conclusion}
\label{sec:conclusion}
The Optimal Transport Gaussian Process (OT-GP) framework is presented, a suite of algorithms that solve the critical wall-time bottleneck and inherent instabilities of GP-accelerated saddle searches. Previous methods, while promising sample-efficiency, introduced
rapidly increasing
computational effort in the hyperparameter optimization that negated practical wall-time gains and suffered from frequent,
failures. The work presented here provides a
diagnosis of these problems and provides a robust, comprehensive solution.

We identified the core of the problem in two areas: a flawed conceptual understanding of the GP's statistical nature and a severe, inherent mathematical fragility in the standard derivative kernel.
Instead of interpreting the hyperparameters as constants with physical meaning,
their function is recognized merely as mathematical tools for model adaptation. This perspective justifies our central innovation: a farthest-point sampling strategy, guided by a chemically aware optimal transport metric, that selects a compact, representative subset of data solely for the hyperparameter optimization step. This
helps deal with the
cubic scaling of the computational effort and results in a halving of the mean wall time compared to its predecessor, while retaining all the data for predictions of new values on the surrogate surface.
There are many more hyperparameter optimization steps than predictions.

Stability follows from three complementary controls. (i) An adaptive logarithmic barrier on \(\sigma_{f}^{2}\) prevents the variance from diverging; (ii) a hyper-parameter-oscillation-detection (HOD) monitor enlarges the FPS subset whenever the optimiser exhibits sign-alternating updates, thereby improving the conditioning of the covariance matrix; and (iii) a data-driven trust-radius, based on the same intensive EMD, automatically limits translation steps to regions supported by the training data. Finally, a
projection removes any
overall
translation or rotation from each trial step, ensuring that the GP kernel sees only genuine internal deformations.

Further state-of-the-art reductions will likely involve symmetry adaption, for example via SOFI \cite{gundeSOFIFindingPoint2024}, or by reinforcement learning of MLIPs driven by foundational frameworks like metatomic and metatensor \cite{bigiMetatensorMetatomicFoundational2025}.

The results from
the
benchmark
calculations
confirm the success of this principled approach. The OTGPD method requires on average less than half the computational effort of the GPDimer method in wall time while preserving the order-of-magnitude reduction in the number of electronic structure calculations. Previous local surrogate methods promised sample efficiency but often failed to deliver practical wall-time performance and robust performance. By confronting and correcting the flawed assumptions and inherent instabilities of these prior approaches, the OTGPD method finally delivers on that promise. The OTGPD requires a median of fewer than 30 electronic structure evaluations and an average time-to-solution under 13 minutes, coupled with a success rate of ninety-six percent, and a performance profile showing that it solves more than 70\% of the benchmark at the best recorded time. The framework developed here provides a powerful approach for active learning, capable of efficiently generating the high-energy, strained transition state geometries essential for training robust, next-generation reactive machine-learned interatomic potentials. The framework’s generality also invites application to entirely new scientific domains, such as complex excited-state potential energy surfaces relevant for photochemistry
and enabling high-throughput adaptive kinetic Monte Carlo simulations coupled directly to electronic structure calculations.
This work elevates GPR acceleration from a promising but computationally heavy tool into a truly robust and wall-time-efficient approach, a foundational technology that enables the routine and automated exploration of complex chemical landscapes.
\section*{Acknowledgments}
\label{sec:acknowledgements}
R.G. received funding from the Icelandic Research Fund (grant number 217436-053) and
financial support from Dr. Guillaume Fraux and Prof. Michele Ceriotti of Lab-COSMO.
The authors thank Prof. Birgir Hrafnkelsson, Prof. Thomas Bligaard, Dr. Andreas Vishart and Dr. Miha Gunde for helpful discussions.
R.G. also
acknowledges valuable discussions with Dr. Amrita Goswami, Dr. Moritz Sallermann, Prof. Debabrata Goswami, Mrs. Sonaly
Goswami, and Mrs. Ruhila Goswami.
The calculations were carried out using resources supplied by the
Icelandic Research e-Infrastructure project
(IREI), supported by the Icelandic Infrastructure Fund.
R.G. presented a preliminary
account of the work presented here at
UNOOS 2025, a conference held in honor of Prof. Debabrata Goswami's on the occasion of his sixtieth birthday.
R.G. warmly dedicates this article to him on this occasion.
\section*{Conflict of Interest}
\label{sec:coi}
We declare no conflicts of interest.

\begin{appendices}
\vspace{0.7em}
\noindent \textbf{Code for reproduction on Github. Data on Materials Archive}
\begin{description}
\item[{Github}] \url{https://github.com/TheochemUI/otgpd\_repro}
\item[{Materials Archive}] \url{https://doi.org/10.24435/materialscloud:rh-tw}
\end{description}
\section{Reproduction note}
\label{sec:org1770dff}
The full set of benchmark inputs, raw outputs, analysis scripts, and pinned runtime environments used in this study are publicly archived. The GitHub repository and Materials Cloud archive contain the original runs for all 238 systems, the scripts used to generate every figure and table, and environment specifications (container/environment manifests) that reproduce the computational environment.

To diagnose a small number of non-convergent cases we performed local re-runs only for the four systems which fail for the OTGPD but not the GPDimer; these runs and their logs are included in the same archive and flagged in the repository for convenience. Users wishing to reproduce any specific experiment can either use the provided raw outputs, the pre-processed FAIR formatted \texttt{csv} data, or re-run the workflow using the included environment manifests and Snakemake pipelines; exact instructions and file paths are given in the archive and on the Github repository.
\section{Rotation removal implementation}
\label{sec:org937de01}
Our Gaussian Process model approximates the potential energy surface without inherent knowledge of the physical invariances of the system. Consequently, a proposed optimization step may contain spurious components corresponding to the external degrees of freedom: overall translation and rotation of the entire molecule. The optimizer actively removes these components from the proposed translation step vector to ensure that movements occur only along internal coordinates, which represent genuine changes in molecular geometry.

The procedure first constructs a basis set spanning the space of infinitesimal rigid-body motions. For a system of \(N\) atoms, this space has six dimensions (or five for a linear molecule). The procedure generates three basis vectors for translation, \(\{\mathbf{t}_x, \mathbf{t}_y, \mathbf{t}_z\}\), where each vector \(\mathbf{t}_k\) represents a unit displacement of all atoms along the Cartesian axis \(k\).

\begin{equation}
(\mathbf{t}_k)_{3i+k-1} = 1 \quad \forall i \in \{1, ..., N\}
\end{equation}

Next, the procedure generates three basis vectors for rotation, \(\{\mathbf{l}_x, \mathbf{l}_y, \mathbf{l}_z\}\), derived from the expression for infinitesimal rotation about the center of mass, \(\mathbf{r}_i' = \mathbf{r}_i - \mathbf{r}_{\text{com}}\). An infinitesimal rotation of the entire system corresponds to a displacement \(\delta\mathbf{r}_i = \delta\boldsymbol{\omega} \times \mathbf{r}_i'\). The rotational basis vectors thus take the form:

\begin{align}
\mathbf{l}_x &= \sum_{i=1}^{N} \hat{\mathbf{e}}_x \times \mathbf{r}_i' \\
\mathbf{l}_y &= \sum_{i=1}^{N} \hat{\mathbf{e}}_y \times \mathbf{r}_i' \\
\mathbf{l}_z &= \sum_{i=1}^{N} \hat{\mathbf{e}}_z \times \mathbf{r}_i'
\end{align}

The algorithm then applies the Gram-Schmidt process to this set of six vectors to produce an orthonormal basis, \(\{\mathbf{u}_k\}\), that spans the external degrees of freedom. For any proposed translation step, \(\mathbf{s} \in \mathbb{R}^{3N}\), the algorithm projects out the external components. The component of the step corresponding to translation and rotation, \(\mathbf{s}_{\text{ext}}\), projects onto this basis:

\begin{equation}
\mathbf{s}_{\text{ext}} = \sum_{k} (\mathbf{s} \cdot \mathbf{u}_k) \mathbf{u}_k
\end{equation}

The pure internal step, \(\mathbf{s}_{\text{int}}\), then becomes the original step minus its external projection:

\begin{equation}
\mathbf{s}_{\text{int}} = \mathbf{s} - \mathbf{s}_{\text{ext}}
\end{equation}

A feedback mechanism enhances the stability of the GP-driven search. The algorithm computes the magnitude of the removed component, \(\|\mathbf{s}_{\text{ext}}\|\). If this magnitude exceeds a defined threshold, \(\theta_{\text{rot}}\), it signals that the GP model likely predicts a large, unphysical torque on the molecule. In such cases, the procedure discards the projection and reverts to the original, unprojected step \(\mathbf{s}\). Subsequent step-size limitation guardrails then typically intercept this large, physically questionable step, triggering a resampling of the true potential energy surface to improve the GP model. When the magnitude of the removed component remains below the threshold, the algorithm accepts the purified internal step \(\mathbf{s}_{\text{int}}\). This ensures a more precise update to the molecular geometry, guided only by genuine internal forces.

In practice, since energy doesn't depend on rotations, the threshold tends to large values.
\section{More computational details}
\label{sec:org0f63ce1}
The analysis was performed on a dataset of 500 initial configurations of small organic molecules with between 7 and 25 atoms.
All versions of software used with the exception of NWChem are vendored.
\begin{itemize}
\item For EON \cite{chillEONSoftwareLong2014}, \href{https://eondocs.org/releases/v2.8.0/}{>2.8.0} should be compatible
\item NWChem \cite{apraNWChemPresentFuture2020} integration requires \href{https://github.com/nwchemgit/nwchem/pull/1145}{this pull request} allowing clients to poll for inputs
\end{itemize}
Searches were aborted if they exceeded 1000 iterations, if the energy increased by more than 20 eV, or if NWChem produced a fatal error. The final convergence criterion for a saddle point was a root-mean-square force below 0.01 eV/Å.
\section{Distance measure quantification}
\label{sec:org226c5e5}
The core of the problem is that the inverse distance we use is not invariant to the permutation of identical atoms, and since kernel's value depends on a direct, index-wise comparison of the interatomic distance vectors of two configurations. This creates a dependency on the arbitrary, fixed labels of the atoms, rather than their physical roles.

An easy way to understand this stems from observing symmetric systems. For instance, consider a proton (indexed k) transferring between two chemically equivalent sites (m and n). Physically, the initial and final states are energetically degenerate. However, a fixed-index comparison metric perceives a significant geometric change, as the distance r(k,m) transitions from short to long, while r(k,n) simultaneously transitions from long to short. The metric fails to recognize that the permutation of labels would reconcile the apparent structural difference.

While the kernel's fitted length-scale hyperparameter may partially average out this effect, a non-averaged metric for early stopping feels the full impact of the flaw. The 1D max log distance, by its definition, registers a significant, non-physical distance for this symmetric swap:

$$D_{\text{1Dmaxlog}}(\mathbf{x}_1, \mathbf{x}_2) = \max_{i,j} \left| \log \frac{r_{ij}(\mathbf{x}_2)}{r_{ij}(\mathbf{x}_1)} \right|$$

\begin{figure}[htbp]
\centering
\includegraphics[width=.9\linewidth]{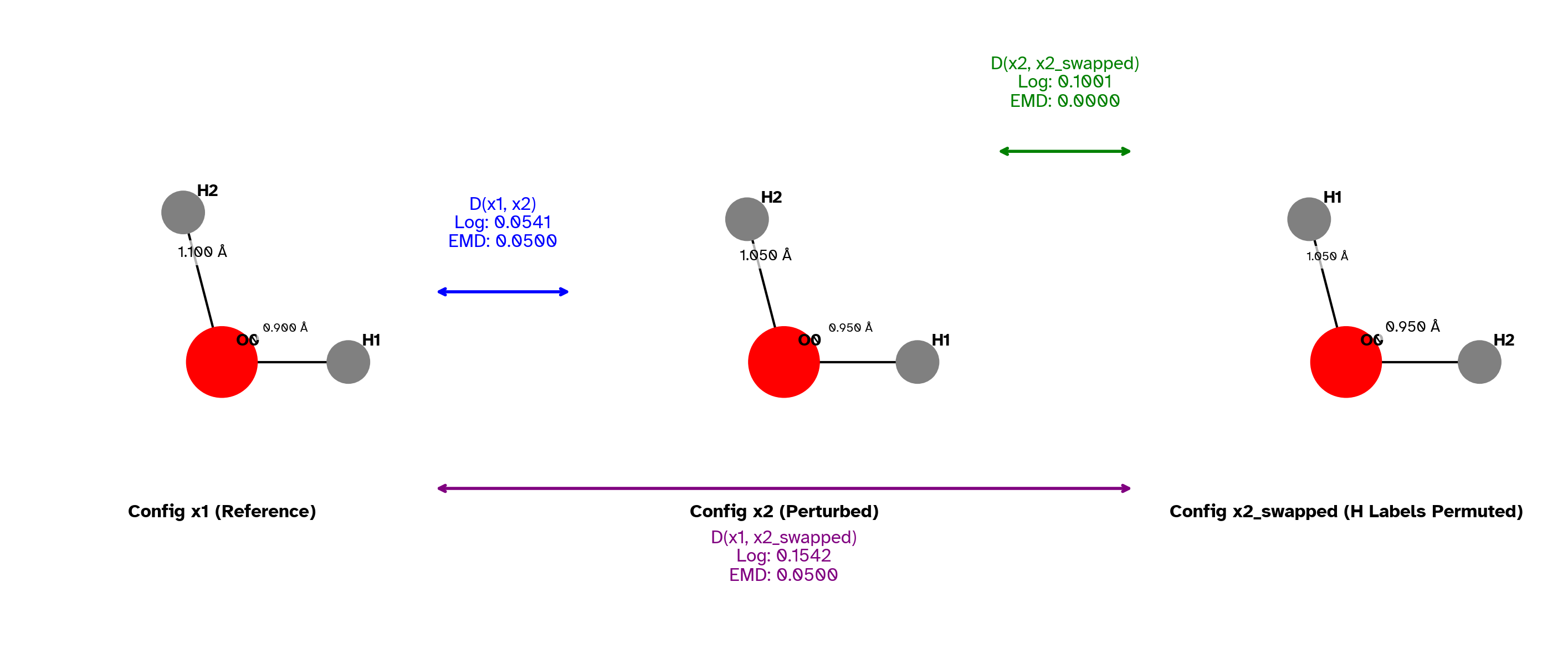}
\caption{\label{fig:suppl:1dmaxEMD}Comparison of the 1D max log distance and the Earth Mover's Distance (EMD) for an asymmetric stretch of a water molecule. While configuration \(x_2\) and \(x_{2,s}\) are physically identical (differing only by the permutation of hydrogen atom labels), the 1D max log metric incorrectly assigns a large distance between them and the reference \(x_1\). In contrast, the EMD correctly identifies them as being equidistant from the reference, demonstrating its permutational invariance.}
\end{figure}

This sensitivity to labeling motivates using the intensive EMD. Figure \ref{fig:suppl:1dmaxEMD} demonstrates this, by contrasting the behavior of both metrics for the asymmetric stretching of a water molecule.

\subsection{Notational clarity}
\label{sec:orga094348}
An inequality expresses the measure in the literature \cite{koistinenMinimumModeSaddle2020}:
$$\frac{2}{3}r_{ij}(\mathbf{x}_{\text{eval}}) < r_{ij}(\mathbf{x}_{\text{im}}) < \frac{3}{2}r_{ij}(\mathbf{x}_{\text{eval}})$$
This is however equivalent.
$$\left| \log \frac{r_{ij}(\mathbf{x}_{\text{im}})}{r_{ij}(\mathbf{x}_{\text{eval}})} \right| < \log(1.5) \approx 0.405$$
\section{Fragment detection}
\label{sec:org1c91b30}
To unambiguously distinguish covalently bonded molecular fragments from transient non-covalent contacts within the simulation cell, a quantum mechanical bonding analysis was performed. For each system configuration, the Mayer-Wiberg-type bond orders for all atom pairs were calculated using the GFN2-xTB semi-empirical method \cite{bannwarthGFN2xTBanAccurateBroadly2019}, as implemented in the \texttt{tblite} library.

Each calculation was performed on the entire supercell. The total charge of the system was set to zero, and the spin multiplicity was explicitly defined as either singlet (1 unpaired electron pair) or doublet (1 unpaired electron), corresponding to the pre-determined electronic state of the system.

From the resulting bond order matrix, a molecular connectivity graph was constructed. A chemical bond between two atoms, i and j, was defined to exist if and only if their calculated bond order, \(BO_ij\), exceeded a threshold, 0.7 here. Based on a calibration procedure for representative configurations, a bond order threshold of 0.7 was selected. This value provides a robust margin to exclude minor, non-zero bond orders (up to \(\approx 0.56\)) calculated for close inter-fragment contacts, while safely including all genuine covalent bonds (typically with BO\textsubscript{ij} > 0.9). Finally, the distinct molecular fragments were identified by determining the connected components of the resulting graph. Fragments with geometric centers less than 1 angstrom away were merged together to take into account slightly fragmented systems.
\section{D136 and the effect of initial conditions}
\label{sec:orgd35dcf2}
This section provides detailed evidence supporting the direct comparison of the
OTGPD and GPDimer methods in the main text. The data demonstrate that the
initial rotation phase on the true potential energy surface effectively
equalizes the starting point for the main search, irrespective of the initial
guess heuristic.
\subsection{Initialization}
\label{sec:org2e5ce69}
The two methods begin with different procedures to generate the initial atomic
displacement.

\begin{description}
\item[{OTGPD}] we implemented a new displacement heuristic in the EON code. This
method displaces the structure by a fixed magnitude of 0.01 Å along the
softest vibrational mode. An inexpensive GFN2-xTB semi-empirical calculation
determines this mode, and the displacement only affects atoms not designated
as frozen.

\item[{GPDimer \cite{goswamiEfficientImplementationGaussian2025a}}] we utilize a
standard displacement heuristic pre-existing in the EON code. This method
follows a multi-step stochastic process: first, it identifies the least
coordinated atom(s) using a 3.3 Å cutoff radius. It then randomly selects one
of these atoms as a displacement epicenter and displaces it along with all
neighboring atoms within a 5.0 Å radius. The displacement for each atom
follows a 3D vector drawn from a Gaussian distribution, where the 0.01 Å
\texttt{displace\_magnitude} parameter serves as the standard deviation.
\end{description}
\subsection{Post-Initialization Comparability for D136}
\label{sec:orgccd5c8a}
We analyze the initial conditions and saddles from a structural perspective
using the Constrained Shortest Distance Assignments (CShDA)
\cite{gundeIRAShapeMatching2021}.

\begin{minted}[]{bash}
cd runs
python init_condcheck.py automated/snake_runs/softest_mode_scg_barrier/doublet/136 \
    ../../gprd_sella_bench/bench_runs/eon/runs/1e8m/gprd/final_gprd_wparam/doublets/136/

                          Structural & Vector Comparison
+------------------------------------+--------------------------------------+
|                             Metric | Value                                |
+------------------------------------+--------------------------------------+
|                    Run 1 Directory | softest_mode_scg_barrier/doublet/136 |
|                    Run 2 Directory | gprd/final_gprd_wparam/doublets/136  |
|------------------------------------+--------------------------------------|
|     Start Structure CSHDA RMSD (Å) | 0.000000                             |
| Displaced Structure CSHDA RMSD (Å) | 0.004925                             |
|    Saddle Structure CSHDA RMSD (Å) | 0.007412                             |
|        Direction Cosine Similarity | 0.008106                             |
+------------------------------------+--------------------------------------+
           INFO     Comparing results.dat metrics...
                          results.dat Comparison
+---------------------------+---------------+---------------+------------+
|                    Metric |         Run 1 |         Run 2 | Difference |
+---------------------------+---------------+---------------+------------+
|          final_eigenvalue |      -42.1910 |      -39.8012 |      +2.39 |
|        force_calls_saddle |            26 |            38 |        +12 |
|                iterations |            23 |            34 |        +11 |
|                  job_type | saddle_search | saddle_search |          - |
| potential_energy_reactant |    -9430.1098 |    -9430.1097 |   +7.7e-05 |
|   potential_energy_saddle |    -9431.1361 |    -9431.1361 |     -2e-06 |
|            potential_type |  SocketNWChem |    ASE_NWCHEM |          - |
|               random_seed |          1995 |     706253457 | +706251462 |
|        termination_reason |       Success |       Success |          - |
|         total_force_calls |            28 |            39 |        +11 |
+---------------------------+---------------+---------------+------------+

Notes:
- CSHDA RMSD closer to 0 means more similar structures (permutation-invariant).
- Cosine Similarity closer to 1 means more similar directions (0=orthogonal, -1=opposite).
\end{minted}

The two heuristics produced nearly orthogonal initial displacement vectors
(cosine similarity \(\approx 0.01\)). However, the subsequent rotation on the true PES, a
process common to both methods, which takes up to 6 iterations on the true
energy surface, effectively erases this initial difference. This equalization
procedure yields nearly identical final saddle structures (RMSD \(\approx 0.007\) Å) and
energies. Consider, also the log for the hyperparameters after the initial
rotation:

\begin{minted}[]{bash}
# GP Dimer
Rotated the initial dimer in 4 outer iterations (total number of image evaluations: 5).
# Hyperparameters
magnSigma2: 0.0304955
lengthScales:
0.102959
0.239712
0.170075
0.235037
0.718695
# Total time
[2025-09-23T09:41:55Z TRACE] pixi r eonclient took 19m 58s 574ms 382us 392ns to complete.
# OTGP-Dimer
Rotated the initial dimer in 2 outer iterations (total number of image evaluations: 3).
# Hyperparameters
magnSigma2: 0.0634641
lengthScales:
0.102861
0.417811
0.161203
 0.58395
0.834778
# Total time
[2025-02-03T03:35:33Z TRACE] pixi r eonclient took 45m 47s 580ms 955us 824ns to complete.
\end{minted}

This show that, the initial hyperparameters learned by each GP model show
significant similarity, as expected for models observing the same local
potential energy surface. The differences stem from SCG barrier method which
changes the likelihood of the OTGP as discussed in the main text. The OTGP-Dimer
method shows larger length scales \((l_2, l_4, l_5)\) compared to GP Dimer. These
larger values indicate that the OTGPD model perceives the potential energy
surface as smoother along these specific coordinates. Collectively, this
demonstrates that the main search phase for both algorithms begins from a
physically equivalent and directly comparable state.
\section{Statistical analysis of cost}
\label{sec:orga13fc80}
Two hierarchical Bayesian regression models were fitted to quantify the impact of the three search algorithms (OT-GPD, GP-Dimer, standard Dimer) on (i) the number of expensive potential-energy-surface (PES) evaluations and (ii) the total wall-clock time required to locate a first-order saddle point.  Both models share the same random-effect structure (a molecule-specific intercept that also accounts for spin multiplicity) and were implemented in the \texttt{brms} R package (v2.19.1) using the \texttt{cmdstanr} backend.

We fit hierarchical Bayesian regression models (negative-binomial for the number of PES evaluations and Gamma for the wall-clock time) with the method (OT-GPD, GP-Dimer, standard Dimer) as a fixed effect and a random intercept for each molecule–spin combination \cite{goswamiBayesianHierarchicalModels2025a}. We report posterior medians together with 95\% credible intervals (CrI) in Tbl. \ref{tbl:mod:perf}.

\begin{table}[htbp]
\caption{\label{tbl:mod:perf}Table of modeled performance metrics}
\centering
\begin{tabular}{lrrl}
Method & Median PES Calls & Time (min) & 95\% CrI Time\\
\hline
OTGPD & 28 & 9.0 & {[}8.6, 9.5]\\
Dimer & 254 & 20.5 & {[}19.2, 21.8]\\
GPDimer & 30 & 12.8 & {[}12.2, 13.5]\\
\end{tabular}
\end{table}

The statistical model confirms the practical gains of the OT-GP framework. Compared with the baseline Dimer, OTGPD achieves an \(\approx\) 89\% reduction in the median number of expensive force evaluations (254 → 28 calls) and more than halves the median time-to-solution (20.5 min → 9.0 min). The improvement over the earlier GP-Dimer also proves decisive: the OTGPD runs \(\approx\) 30\% faster (9.0 min versus 12.8 min) while using slightly fewer force calls. The posterior distributions for these effects show clear separation from zero, indicating a >99\% probability that OTGPD outperforms both comparators.

We may use such models for interpretation, understanding that these provide additional insights into the distributional details of the raw data.
\subsection{Model specifications}
\label{sec:orge1b1c3f}
\subsubsection{Prior distributions}
\label{sec:org4c5a7b0}
We use weakly informative priors, chosen to regularise the models without imposing strong beliefs about the magnitudes of effects, detailed in Table \ref{tbl:suppl:priorpred}.

\begin{table}[htbp]
\caption{\label{tbl:suppl:priorpred}Priors for predictive model}
\centering
\begin{tabularx}{\textwidth}{|X|l|X|}
Parameter & Prior (distribution) & Rationale\\
\hline
Fixed-effect coefficients (\(\beta_{method}\), \(\gamma_{method}\)) & \texttt{normal(0, 1)} & Allows a 95 percent prior range of roughly \textpm{} 2 on the log scale (\(\approx e^{-2}\) to e² \(\approx\) 0.14–7.4), comfortably covering plausible speed-up or slow-down factors.\\
Intercept (\(\beta_o\), \(\gamma_o\)) & \texttt{student\_t(3, 0, 2.5)} & Heavy-tailed to tolerate outliers while still centring near zero.\\
Random-effect standard deviation (\(\sigma_mol\)) & \texttt{exponential(1)} & Strongly shrinks the group variance toward zero unless data demand otherwise.\\
Shape parameter of the negative-binomial (\(\Psi\)) & \texttt{exponential(1)} & Enforces positivity and encourages modest over-dispersion.\\
Shape parameter of the Gamma (\(\alpha\)) & \texttt{exponential(1)} & Guarantees positivity and avoids overly heavy tails.\\
\end{tabularx}
\end{table}

We specified priors identically for both models, the \texttt{exponential(1)} prior for scale/shape parameters has a mean of 1, reflecting a modest expectation of dispersion.
\subsubsection{Computational settings}
\label{sec:orgb67588b}
Both models were fitted with four parallel Markov chains, each with 4 000 iterations (the first 1 000 discarded as warm-up). The total effective sample size for all parameters exceeded 1 200 (\(\geq 3 \times\) the number of chains) and the potential scale reduction factor (\(\hat R\)) was \(\leq\) 1.01 for every parameter, indicating convergence. Sampling was performed with the NUTS (No-U-Turn Sampler) algorithm implemented in CmdStan (v2.35.0) via the \texttt{cmdstanr} R interface. To ensure stable Hamiltonian dynamics we set \texttt{adapt\_delta = 0.99} and \texttt{max\_treedepth = 15} for the wall-time model (the negative-binomial model converged with the default settings).
\subsubsection{Number of PES calls}
\label{sec:org0191c3c}

The count data may be over-dispersed relative to a Poisson distribution, therefore we employ a negative-binomial likelihood (log link):

\[
\begin{aligned}
\text{pes\_calls}_{i} &\sim \operatorname{NegBinomial}(\mu_{i},\phi) \\
\log(\mu_{i}) &= \beta_{0} + \beta_{\text{method}[i]} + u_{j[i]} \\
u_{j} &\sim \mathcal{N}(0,\sigma_{\text{mol}}^{2}) \quad (j = \text{mol\_id:spin}) .
\end{aligned}
\]

\begin{itemize}
\item \emph{Fixed effects} (\(\beta_{method}\)) encode the multiplicative shift relative to the baseline Dimer (reference level).
\item \emph{Random intercept} (\(u_j\)) captures systematic differences among molecules and between spin states.
\end{itemize}

\begin{minted}[]{r}
 Family: negbinomial
  Links: mu = log
Formula: pes_calls ~ method + (1 | mol_id:spin)
   Data: data (Number of observations: 692)
  Draws: 4 chains, each with iter = 4000; warmup = 1000; thin = 1;
         total post-warmup draws = 12000

Multilevel Hyperparameters:
~mol_id:spin (Number of levels: 238)
              Estimate Est.Error l-95% CI u-95% CI Rhat Bulk_ESS Tail_ESS
sd(Intercept)     0.35      0.02     0.31     0.40 1.00     4042     7082

Regression Coefficients:
              Estimate Est.Error l-95% CI u-95% CI Rhat Bulk_ESS Tail_ESS
Intercept         5.61      0.03     5.55     5.67 1.00     6816     7994
methodGPDimer    -2.16      0.03    -2.22    -2.10 1.00    17086     9663
methodOTGPD      -2.26      0.03    -2.32    -2.20 1.00    16874     9285

Further Distributional Parameters:
      Estimate Est.Error l-95% CI u-95% CI Rhat Bulk_ESS Tail_ESS
shape    11.26      0.87     9.65    13.03 1.00     7688     8984

Draws were sampled using sample(hmc). For each parameter, Bulk_ESS
and Tail_ESS are effective sample size measures, and Rhat is the potential
scale reduction factor on split chains (at convergence, Rhat = 1).
\end{minted}

With the exact results in Table \ref{tbl:suppl:pes}.

\begin{table}[htbp]
\caption{\label{tbl:suppl:pes}Results of the PES model}
\centering
\begin{tabular}{lrl}
Effect\textsubscript{Type} & Median Effect & 95\% CrI\\
\hline
Expected PES Calls (Baseline: Dimer) & 272.8 & {[}256.7, 289.9]\\
Multiplicative Factor (GPDimer vs Dimer) & 0.1 & {[}0.11, 0.12]\\
Percentage Change (GPDimer vs Dimer) & -88.4\% & {[}-89.1\%, -87.7\%]\\
Multiplicative Factor (OTGPD vs Dimer) & 0.1 & {[}0.10, 0.11]\\
Percentage Change (OTGPD vs Dimer) & -89.6\% & {[}-90.2\%, -88.9\%]\\
sd(Intercept) [mol\textsubscript{id}:spin] & 0.3 & {[}0.31, 0.40]\\
\end{tabular}
\end{table}
\subsubsection{Total wall-clock time}
\label{sec:org424c82e}

Since wall-clock times follow a continuous, strictly positive, and right-skewed distribution, we use a Gamma likelihood with a log link. To capture the complex, non-linear relationship between the number of PES evaluations and the total time, which can differ significantly between methods due to varying overhead costs, we employ a generalized additive model (GAM). Specifically, we model the log of the expected time as a function of method-specific smoothing splines of the log-transformed PES calls:

\[
\begin{aligned}
\text{tot\_time}_{i} &\sim \operatorname{Gamma}(\alpha,\beta_{i}) \\
\log(\beta_{i}) &= \gamma_{0} + \gamma_{\text{method}[i]} + s_{\text{method}[i]}(\log(\text{pes\_calls}_{i})) + v_{j[i]} \\
v_{j} &\sim \mathcal{N}(0,\sigma_{\text{mol}}^{2}) .
\end{aligned}
\]

In this formulation, the term \(s_{\text{method}[i]}(\cdot)\) represents a unique thin-plate regression spline for each method, allowing the model to learn the distinct, non-linear time-cost profiles. The fixed effects (\(\gamma_{\text{method}}\)) capture the baseline differences between methods, while the random intercept (\(v_j\)) accounts for system-specific variations, as in the PES calls model. This more flexible structure allows for a more accurate and nuanced comparison of method efficiencies across their operational ranges.

\begin{minted}[]{r}
 Family: gamma
  Links: mu = log
Formula: tot_time ~ method + s(log_pes_calls, by = method, k = 5) + (1 | mol_id:spin)
   Data: data (Number of observations: 1433)
  Draws: 4 chains, each with iter = 4000; warmup = 1000; thin = 1;
         total post-warmup draws = 12000

Smoothing Spline Hyperparameters:
                                   Estimate
sds(slog_pes_callsmethodDimer_1)       2.44
sds(slog_pes_callsmethodGPDimer_1)     3.79
sds(slog_pes_callsmethodOTGPD_1)       4.15
                                   Est.Error
sds(slog_pes_callsmethodDimer_1)        0.51
sds(slog_pes_callsmethodGPDimer_1)      0.52
sds(slog_pes_callsmethodOTGPD_1)        0.51
                                   l-95% CI
sds(slog_pes_callsmethodDimer_1)       1.56
sds(slog_pes_callsmethodGPDimer_1)     2.87
sds(slog_pes_callsmethodOTGPD_1)       3.22
                                   u-95% CI Rhat
sds(slog_pes_callsmethodDimer_1)       3.53 1.00
sds(slog_pes_callsmethodGPDimer_1)     4.89 1.00
sds(slog_pes_callsmethodOTGPD_1)       5.24 1.00
                                   Bulk_ESS
sds(slog_pes_callsmethodDimer_1)       8533
sds(slog_pes_callsmethodGPDimer_1)    10689
sds(slog_pes_callsmethodOTGPD_1)      12370
                                   Tail_ESS
sds(slog_pes_callsmethodDimer_1)       6729
sds(slog_pes_callsmethodGPDimer_1)     7913
sds(slog_pes_callsmethodOTGPD_1)       8356

Multilevel Hyperparameters:
~mol_id:spin (Number of levels: 499)
              Estimate Est.Error l-95% CI u-95% CI
sd(Intercept)     0.39      0.02     0.35     0.42
              Rhat Bulk_ESS Tail_ESS
sd(Intercept) 1.00     3036     6241

Regression Coefficients:
                               Estimate Est.Error
Intercept                          2.71      0.12
methodGPDimer                      1.15      0.22
methodOTGPD                       -0.11      0.28
slog_pes_calls:methodDimer_1       1.51      0.95
slog_pes_calls:methodGPDimer_1     0.44      1.00
slog_pes_calls:methodOTGPD_1       0.22      1.00
                               l-95% CI u-95% CI
Intercept                          2.48     2.94
methodGPDimer                      0.72     1.58
methodOTGPD                       -0.67     0.44
slog_pes_calls:methodDimer_1      -0.33     3.38
slog_pes_calls:methodGPDimer_1    -1.54     2.40
slog_pes_calls:methodOTGPD_1      -1.74     2.16
                               Rhat Bulk_ESS
Intercept                      1.00    10213
methodGPDimer                  1.00     7315
methodOTGPD                    1.00     7428
slog_pes_calls:methodDimer_1   1.00    13657
slog_pes_calls:methodGPDimer_1 1.00    19482
slog_pes_calls:methodOTGPD_1   1.00    19934
                               Tail_ESS
Intercept                          9335
methodGPDimer                      8663
methodOTGPD                        7717
slog_pes_calls:methodDimer_1       9152
slog_pes_calls:methodGPDimer_1     8434
slog_pes_calls:methodOTGPD_1       9243

Further Distributional Parameters:
      Estimate Est.Error l-95% CI u-95% CI Rhat
shape     7.73      0.36     7.04     8.47 1.00
      Bulk_ESS Tail_ESS
shape     6529     8211

Draws were sampled using sample(hmc). For each parameter, Bulk_ESS
and Tail_ESS are effective sample size measures, and Rhat is the potential
scale reduction factor on split chains (at convergence, Rhat = 1).
\end{minted}

\begin{figure}[htbp]
\centering
\includegraphics[width=.9\linewidth]{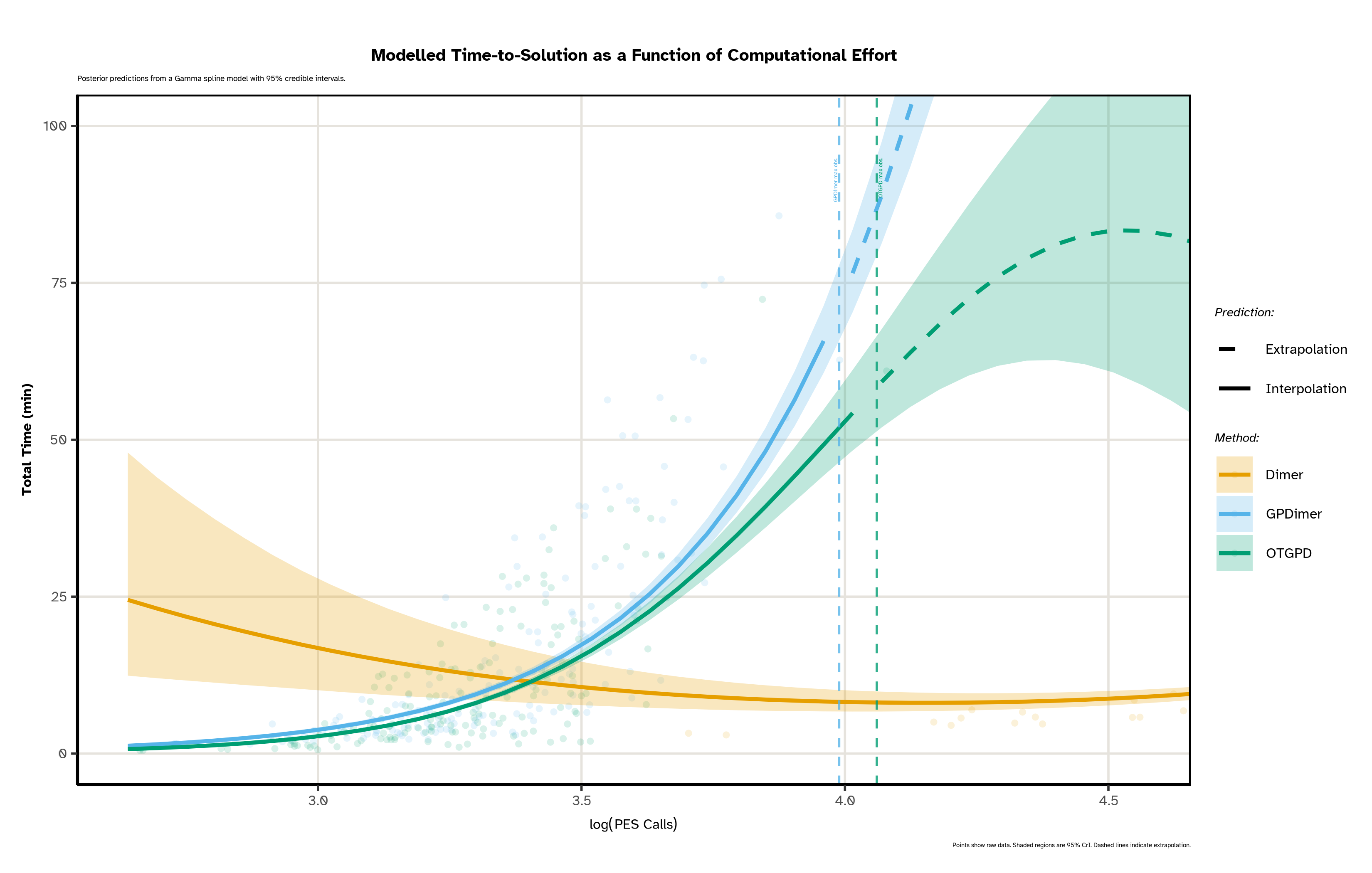}
\caption{\label{fig:suppl:wtimepes}Posterior predictions from the hierarchical Gamma spline model for total wall-clock time as a function of PES evaluations. Points represent the raw data for each of the three methods. Solid lines are the model's posterior mean predictions, with shaded regions indicating the 95\% credible intervals. The relationship between time and PES calls is allowed to be non-linear and is modeled independently for each method. For the GP-based methods, dashed lines indicate where the model is extrapolating beyond the maximum number of observed PES calls for that method, which are marked by the vertical dashed lines.}
\end{figure}

We show the model predictions overlaid with the data in Figure \ref{fig:suppl:wtimepes}.

While this model provides a robust estimate of the median performance over the data seen, the full extent of OTGPD's superiority, particularly its ability to avoid the worst-case timings that affect GPDimer, is most clearly visualized by the performance profiles in the main text. The model shows that OTGPD reduces the median time-to-solution to less than a half compared to the dimer method (9.0 min vs 20.5 min) and runs \(\approx\) 30\% faster than the GPDimer (9.0 min vs 12.8 min), with a greater than 99\% probability that OTGPD outperforms the other two methods.
\section{GPDimer / OTGPD failure modes}
\label{sec:org4cadd5d}

The OTGPD framework successfully eliminates the signal variance instability. 4 failures were observed in the OTGPD benchmark but not in the GPDimer which we consider in this section. These were not due to an algorithmic instability but were artifacts of the dimer initialization routine.

For the OTGPD benchmark, initial dimer configurations were generated by displacing atoms along the softest mode found by an inexpensive xtb calculation. For systems D016, D084, D100, and S242, this procedure created a pathological starting geometry with excessively high forces or atoms in unphysically close contact.

\begin{figure}[htbp]
\centering
\includegraphics[width=.9\linewidth]{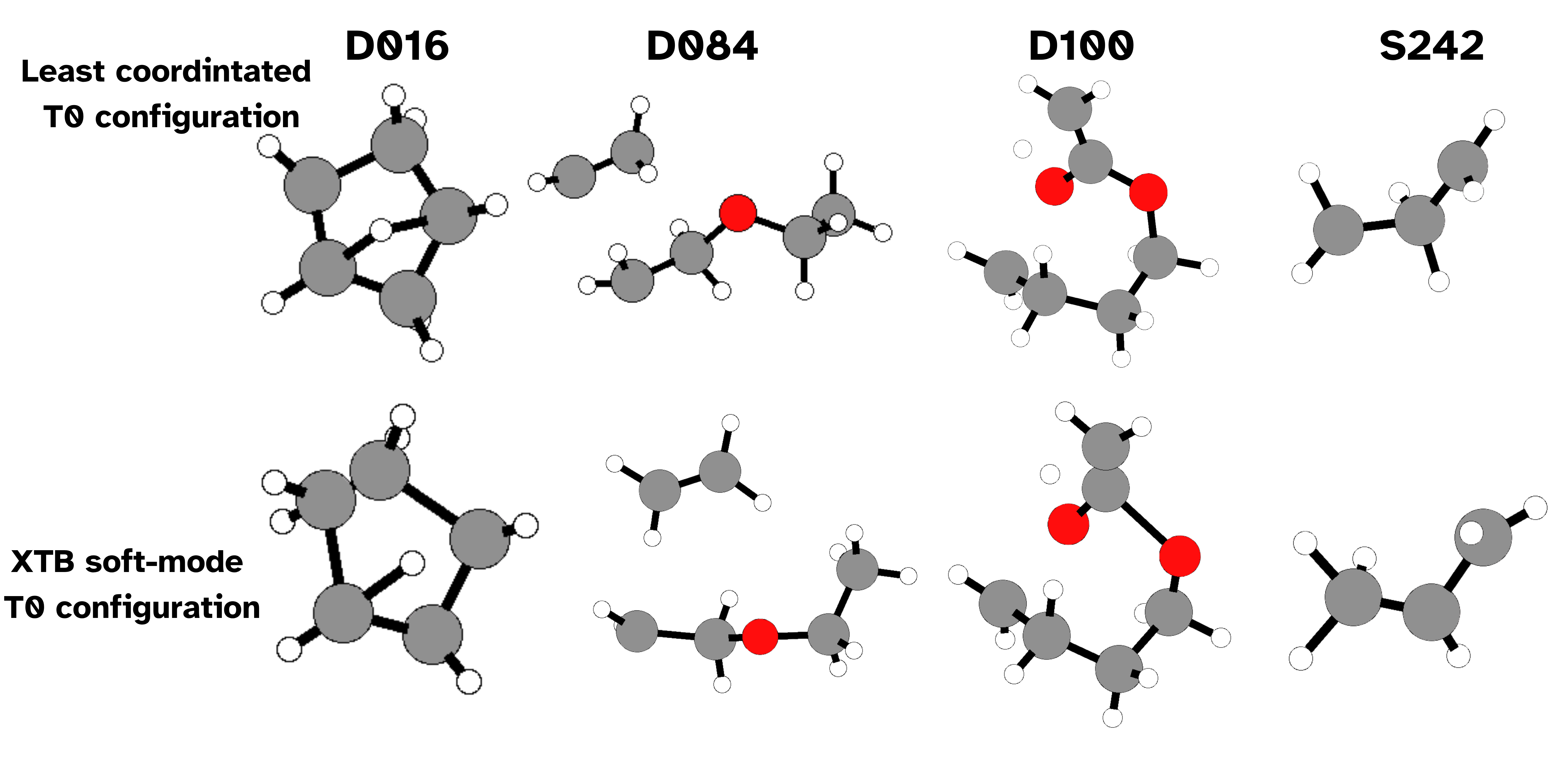}
\caption{\label{fig:suppl:initgpdotgpd}Initializations for the GPDimer (top) and OTGPD (bottom). The XTB initialization procedure in these four cases results in starting the runs from unphysical initial geometries leading to failures in the optimization routine. With XTB, D016 has unphysically close carbon atoms, D084 has a shortened carbon-oxygen bond, D100 has near-overlapping carbon atoms, and S242 has a CH3 end-group instead of the hydrogen on the middle carbon.}
\end{figure}

When any GP-accelerated method is ``poisoned'' with such a high-energy, high-force baseline, the initial surrogate model learns that these unphysical configurations are normal, leading to an unstable search that fails to converge. The GPDimer benchmark, used a different (and in these cases, fortuitously non-pathological) initialization based on displacing the least coordinated atom, bypassed this specific problem. Figure \ref{fig:suppl:initgpdotgpd} shows the initial points at the start of the algorithm runs, extracted from the trajectory files.

To confirm this diagnosis, we re-ran these four systems using OTGPD but starting
from the exact same (valid) initial configurations used by the GPDimer runs to show all methods converge.

Times reported in minutes --> \texttt{total seconds / 60}. Calls represent the number of Hartree-Fock calculations, which are the ``true energy samples'' required.
\subsection{D016}
\label{sec:org9c66cf5}

\begin{center}
\begin{tabular}{lrr}
Method & Calls & Time\\
\hline
Dimer & 106 & 7.4\\
GPDimer & 20 & 3.7\\
OTGPD & 20 & 3.3\\
\end{tabular}
\end{center}
\subsection{D084}
\label{sec:orge852b6f}

\begin{center}
\begin{tabular}{lrr}
Method & Calls & Time\\
\hline
Dimer & 2666 & 255.4\\
GPDimer & 75 & 417\\
OTGPD & 65 & 181.4\\
\end{tabular}
\end{center}
\subsection{D100}
\label{sec:org6de2ce4}

\begin{center}
\begin{tabular}{lrr}
Method & Calls & Time\\
\hline
Dimer & 214 & 24.3\\
GPDimer & 28 & 16.2\\
OTGPD & 28 & 18.52\\
\end{tabular}
\end{center}
\subsection{S242}
\label{sec:orgc6c3556}

\begin{center}
\begin{tabular}{lrr}
Method & Calls & Time\\
\hline
Dimer & 249 & 14.6\\
GPDimer & 25 & 2.3\\
OTGPD & 33 & 1.85\\
\end{tabular}
\end{center}
\setlength{\bibsep}{0.0cm}
\clearpage
\end{appendices}
\bibliography{emdGPRD,manual_refs}
\end{document}